\documentstyle[aps,manuscript,epsf]{revtex}

\begin{document}

\begin{center}{\Large \bf Radion and Large Scale Anisotropy on the Brane
} \\
\vskip.25in
{Kazuya Koyama}

{\it $^1$ Department of Physics, The University of Tokyo, 
       Tokyo 113-0033, Japan  \\} %

\end{center}
\begin{abstract}                
We investigate the effect of the radion on cosmological 
perturbations in the brane world. 
The $S^1/Z_2$ compactified 5D Anti-de Sitter
spacetime bounded by positive and negative tension branes is considered.
The radion is the relative displacement of the branes in this model. 
We find two different kinds of the radion at the linear perturbation order
for a cosmological brane. 
One describes a "fluctuation" of the brane which does not couple
to matter on the brane. The other describes 
a "bend" of the brane which couples to the matter. The bend 
determines the curvature perturbation on the brane.
At large scales, the radion interacts 
with anisotopic perturbations in the bulk. 
By solving the bulk anisotropic perturbations, 
large scale metric perturbations and anisotropies of the 
Cosmic Microwave Background (CMB) on the positive tension brane are 
calculated. We find an interesting fact that the radion contributes to
the CMB anisotropies. The observational consequences of these 
effects are discussed. 
\end{abstract}

\newpage

\section{Introduction}
The string theory suggests the idea of 
confining the standard model particles in a 3-brane in a higher dimensional 
spacetime. Based on this brane world idea, Randall and Sundrum proposed
a very interesting scenario that we are living on either of two boundary
3-branes in a 5D Anti-de Sitter (AdS) spacetime \cite{RS1,RS2}.
One of the fascinating features of their scenario is that
the gravity of the positive tension brane is confined near the barne
even if an extra dimension is infinitely large \cite{RS2}. 
Because their scenario provides a new picture for our universe, it is 
important to test their scenario from the cosmological 
view point. For this purpose, the behaviour of the cosmological 
perturbations has been actively investigated [3-15]. 

In this paper, we shall investigate a cosmological implication of a new 
geometrical degree of freedom introduced in their model, that is, 
the displacement of the brane in an extra-dimension. 
We consider branes which are located at $Z_2$ symmetric orbifold 
fixed points. If we take an appropriate coordinate gauge, the displacement 
of the brane is described by the scalar field living on the brane. 
Then, the displacement of the Minkowski brane $\varphi$ 
obeys the equation
\begin{equation}
\Box_4 \varphi = \frac{\kappa^2}{6} T, 
\label{0-1}
\end{equation}
where $T$ is the trace of the energy-momentum tensor of the matter
on the brane and $\Box_4$ is the d'Alembertian operator in 
4D Minkowski spacetime \cite{GT}.  Thus there are two kinds of the displacement
of the brane according to two kinds of the solutions
for Eq. (\ref{0-1}), i.e. the homogeneous and the particular
solutions. The homogeneous solution for 
Eq. (\ref{0-1}) with $T=0$ represents a "fluctuation" of the brane.
The brane can fluctuate by itself without the matter energy-momentum tensor. 
The other kind of the displacement is a "bend" of the brane
due to the matter on the brane. 
The trace of the energy-momentum tensor acts as an tension, then
the brane bends due to this effective tension. 

Of particular interest is the detectability of the brane fluctuation.
A similar situation was considered in the analysis of 
the fluctuation of a thin domain wall \cite{VB}. It was shown that
the wall fluctuation cannot be seen by an interior observer
on the wall because the fluctuation does not change the curvature of the 
domain wall. However this analysis is only done for a test domain wall. 
Once we take into account the gravitational perturbations, we may 
see the effect of the fluctuation \cite{II}. 

The same observation applies to the brane fluctuation. 
If there is only one brane \cite{RS2}, 
the bulk has a translational invariance, at 
least when the energy density of the matter on the brane is 
sufficiently lower than the tension of the brane. Then the gravitational
perturbations are not affected by the brane fluctuation. 
Because the brane fluctuation can be detected only through the 
interaction with gravitational perturbations in the bulk, 
the brane fluctuation has no physical degree of freedom in one brane 
model. \footnote{We should note that $Z_2$ symmetry is an essential 
assumption to obtain this result. If there is no $Z_2$ symmetry then 
there is an extra degree of freedom for brane fluctuations.}
However, if we consider two branes model with $S^1/Z_2$ 
orbifold extra dimension \cite{RS1}, gravitational perturbations 
are confined between two branes.  The displacements
of the branes change the distance between two branes,
then they should affect the gravitational perturbations. Thus
the brane fluctuation acquires a physical degree of the freedom 
and we find a possibility to detect it.
The brane fluctuation changes the size of the extra dimension.
Hence, the brane fluctuation is called radion in literature.
The role of the radion was studied for the Minkowski brane by 
Charmousis, Gregory and Rubakov \cite{CGR}. 
The radion on the de Sitter brane 
was studied classically \cite{US} and quantum mechanically 
\cite{US2} by Gen and Sasaki. 
Chacko and Fox also studied the radion on the AdS brane \cite{CF}. 
The dynamics of the homogeneous radion for an expanding brane was
studied by Bin{\'e}truy, Deffayet and Longlois \cite{BDL}.
There are also several works whcih study the radion 
using the 4D effective action
\cite{action}. 

It was shown that, in two branes model, the gravity on the branes 
is described by the Brans-Dicke (BD) theory where the radion acts as 
a BD scalar \cite{GT}. 
If we are living on a positive tension brane, a BD parameter 
can be compatible with observations if the distance between two branes 
is sufficiently large. 

Now let us consider a cosmology based on this scenario.
The branes expand due to the matter energy momentum tensor on the brane.
Interestingly, the dynamics of a homogeneous and isotropic brane is 
determined only by the matter energy-momentum tensor on the brane
if the bulk is purely AdS spacetime. Thus the homogeneous radion
does not affect the evolution of the homogeneous and isotropic brane. 
However, if one allows the spatial anisotropy of the brane, 
the situation changes significantly. 
The anisotropic perturbation in the bulk is allowed to be exist. 
Then the homogeneous radion interacts with the anisotropic 
bulk perturbation and contributes to the anisotropy of the positive tension 
brane. Our real universe has an anisotropy which is measured by the temperature 
anisotropy of the Cosmic Microwave Background (CMB). Thus it is important to
clarify the contribution of the homogeneous radion on CMB anisotropies.

If there are matter perturbations on the brane, the trace of their
energy momentum tensor inevitably bends the brane \cite{GT}. 
The bend of the brane affects the intrinsic curvature of the brane. 
Thus an understanding of the dynamics of the bend is also
important to know the behavior of the cosmological perturbations on
the brane.
It should be emphasized that the curvature perturbation 
on uniform density hypersurface is determined independently of 
bulk perturbations at large scales. 
This independence of the curvature perturbation from
the bulk perturbation is the important point to understand 
the large scale cosmological perturbations.
This result should be related to the dynamics of the 
bend of the brane.

Although the curvature perturbation can be determined independently 
of the bulk perturbation, CMB anisotropies cannot be predicted
solely by the curvature perturbation. This is because the anisotropic 
perturbation in the bulk is induced by the homogeneous displacement of the brane. 
It also contributes to CMB anisotropies. Then in order to find CMB
anisotropies, we should solve the evolution equation for the bulk perturbation. 
It is another important point to understand the cosmological perturbations
that the anisotropy of the brane cannot be calculated unless the bulk 
perturbation is known. 

Unfortunately, it is a very difficult
task to solve the equation exactly. Thus in this paper, we assume
that the system is nearly static when we solve the bulk equation.
This assumption is valid under two conditions. First, the energy density 
of the matter on the brane is sufficiently lower than the tension 
of the brane. Secondly, the distance between two branes changes 
with time very slowly. Under this assumption, we can solve the 
bulk anisotropic perturbation. Then, we can evaluate large scale 
CMB anisotropies on the positive tension brane and discuss the 
cosmological implication of the radion.

The structure of this paper is as follows. In section II, we describe the
set up of our model. In section III, the formulation to calculate the 
metric perturbations on the brane are shown. The brane displacement
is defined at the linear perturbation order and its coupling to  
bulk perturbations are studied. 
In section IV, the dynamics of the 
brane displacement is investigated. In section V, the bulk anisotropic
perturbation induced by brane displacement is solved. 
In section VI, the solutions for metric perturbations
on a positive tension brane is obtained. The large scale CMB anisotropy
is derived. In section VII we summarize our results. 

\section{Set up of the model}
We consider two branes which are sitting at the 
$S_1/Z_2$ orbifold fixed points in 5D AdS spacetime. 
Our system is described by the action
\begin{equation}
S= \frac{1}{2 \kappa^2}\int d^5 x \sqrt{-g}
\left(
{\cal R}^5 +  \frac{12}{l^2} \right)
+\sum_{i=A,B} \left(- \mu^i \int d^4 x \sqrt{-g_{brane \: i}}
+ \int d^4 x \sqrt{-g_{brane \: i}} {\cal L}_{matter^i}
\right),
\end{equation}
where ${\cal R}^5$ is the 5D Ricci scalar, $l$ is the 
curvature radius of the AdS spacetime and $\kappa^2=8 \pi G_5$ 
where $G_5$ is the 5D Newton constant. 
We assumed that the brane A has the positive tension 
$\mu^A$ and the brane B has the negative tension $\mu^B$ 
respectively. Those tensions are taken as 
\begin{equation}
\kappa^2 \mu^A = \frac{6}{l} ,\quad
\kappa^2 \mu^B = -\frac{6}{l},
\end{equation}
in order to ensure that the branes become
Minkowski spacetime without matter on the brane. 
The induced metric on the brane $i$, either A or B,  
is denoted as $g_{brane \: i}$ and the matter which is confined in the 
brane $i$ is described by the Lagrangian ${\cal L}_{matter^i}$.
We assume that we are living on the positive tension brane A and 
we do not write explicitly the index $i=A$ in the following contents. 

We adopt a coordinate system where both 
branes sit at fixed values of the extra coordinate 
because the physics of the radion is clear in such a basis [19-22].
(For an alternative approach which employs the coordinate system 
where one of the brane is at rest see \cite{BDL}.)
The metric in this coordinate system is taken as 
\begin{equation}
ds^2=e^{2 \gamma(y,t)}dy^2- e^{2 \beta(y,t)} dt^2+ e^{2 \alpha(y,t)}
\delta_{ij} dx^i dx^j.
\label{2-1-1}
\end{equation}
The extra coordinate $y$ is compact and runs from $-l$ to
$l$. Furthermore, the identification of $(y,t,x^i)$ with
$(-y,t,x^i)$ is made. Then the extra dimension becomes $S^1/Z_2$
orbifold space. 
The brane A is located at $y=0$ and the brane B is located at $y=l$ 
respectively. 
The radion ${\cal R}$ is defined as the physical distance between 
two branes [19-23]. Then the radion in the background spacetime ${\cal R}_0$
is given by  
\begin{equation}
{\cal R}_0(t) = \int^{l}_0 dy \: e^{\gamma(y,t)}.
\end{equation}

The 5D energy momentum 
tensor (including the tensions of the branes) is taken as 
\begin{eqnarray}
T^M_N &=& 
\left[ \left(-\frac{6}{\kappa^2 l} \mbox{diag}(0,1,1,1,1)+
\mbox{diag}(0,-\rho,p,p,p) \right)\delta(y) \right. \nonumber\\
&+& \left. \left(\frac{6}{\kappa^2 l} \mbox{diag} (0,1,1,1,1)+
\mbox{diag}(0,-\rho^B,p^B,p^B,p^B) \right)\delta(y-l) \right].
\label{2-1-2}
\end{eqnarray}
In the following paper, we will denote the power series expansion of 
some function $F(y,t)$ around $y=0$ as  
\begin{equation}
F(y,t) = F_{0} (t)+ F_{1}(t) \vert y \vert  + 
\frac{F_{2}(t)}{2} y^2+ 
\cdot \cdot \cdot. 
\end{equation}
In a similar way, the power series expansion of the same 
function $F(y,t)$ around $y=l$ is denoted as 
\begin{equation}
F(y,t) = F_{0}^B(t)+ F_{1}^B(t) \vert y-l \vert  + 
\frac{F_{2}^B(t)}{2} \vert y-l \vert^2+ 
\cdot \cdot \cdot. 
\label{2-1-3}
\end{equation}
5D Einstein equations give the evolution equations 
for the metric $\alpha$, $\beta$ and $\gamma$. Any set of functions
$\alpha$, $\beta$ and $\gamma$ are the solutions for all components 
of 5D Einstein equations when they satisfy the equations 
\begin{eqnarray}
e^{-2 \gamma} \alpha'^2-e^{-2 \beta} \dot{\alpha}^2 &=& \frac{1}{l^2}
-C e^{-4 \alpha}, \nonumber\\
\beta' \dot{\alpha}-\dot{\alpha}'-\dot{\alpha} \alpha'+ \alpha' 
\dot{\gamma} &=& 0,
\label{2-1-4}
\end{eqnarray}
where prime denotes the derivative with respect to $y$ and dot denotes 
the derivative with respect to $t$ \cite{BDEL}
(see Appendix A).
Here $C$ is the integration constant which is related to the mass of the 
black hole in the bulk. The first derivatives of the metric with respect to 
$y$ evaluated at the brane A, i.e. $\alpha_1$ and $\beta_1$, are related to the 
energy-momentum tensor of the matter on the brane by the junction conditions;
\begin{eqnarray}
\alpha_{1}(t) &=& - e^{\gamma_{0}} \left(\frac{1}{l}   
+ \frac{\kappa^2 \rho(t)}{6} \right), \nonumber\\
\beta_{1}(t) &=&  - e^{\gamma_{0}} \left( \frac{1}{l}-
\kappa^2 \rho(t) \left(\frac{1}{3} + \frac{w}{2} \right) \right),
\label{2-1-5}
\end{eqnarray}
where we write $p=w \rho$.
The evolution equations for the energy density $\rho$ and the scale factor
$\alpha_0$ are obtained from the $y^0$-th order of the power series expansion 
of Eq. (\ref{2-1-4});
\begin{eqnarray}
e^{-2 \gamma_0} \alpha_1^2-e^{-2 \beta_0} \dot{\alpha}_0^2 &=&
\frac{1}{l^2}-C e^{-4 \alpha_0}, \nonumber\\
\dot{\alpha_1}-\alpha_1 \dot{\gamma}_0 
-(\beta_1-\alpha_1) \dot{\alpha}_0 &=& 0.
\label{2-1-6-0}
\end{eqnarray}
Then using the junction conditions (\ref{2-1-5}), the evolution equations
for $\alpha_0$ and $\rho$ are obtained as 
\begin{eqnarray}
\dot{\alpha}_0^2 &=& e^{2 \beta_0} \left( \frac{\kappa^2}{3 l} \rho
+\frac{\kappa^4}{36} \rho^2  +C e^{-4 \alpha_0} \right), \nonumber\\
\dot{\rho} &+& 3(1+w) \dot{\alpha}_{0} \rho=0.
\label{2-1-6}
\end{eqnarray}
In this paper, we assume that the bulk is a purely AdS 
spacetime. Then we take the integration constant as $C=0$. 
The interesting point is that the
evolution of the brane universe can be determined locally. 
We do not need to find the solutions for the whole 5D spacetime. 
The radion ${\cal R}_0(t)$ does not play a role to determine 
the evolution of the branes.

When we consider the perturbations, the explicit form of the 
bulk metric should be known. The general solution for the bulk
has been found \cite{exact}. However, as mentioned above, we want to
find a particular solution where both branes sit at fixed
values of the extra coordinate. Such a particular solution for 
the bulk metric was found only for the matter 
energy momentum tensor on the branes with some specific form \cite{BDL}.
In addition, we also need to solve the bulk evolution equation for 
perturbations. It is generally difficult to solve the evolution
equations in the bulk exactly. Hence, we will solve the evolution
equation in the bulk by assuming that the system is nearly static. 
To do so, we assume 
the energy density of the matter on the brane is sufficiently lower than
the tension of the brane
\begin{equation}
\kappa^2 l \rho \ll 1, \quad \kappa^2 l \rho^B \ll 1.
\end{equation}
From the Friedmann equation (\ref{2-1-6}), the time derivative of the metric 
 $\dot{\alpha}$ is of the order $(\kappa^2 l^{-1} \rho)^{1/2}$.
On the other hand, from the junction condition (\ref{2-1-5}), the 
$y$ derivative of the metric $\alpha'$ is of the order $l^{-1}$ 
for $\kappa^2 l \rho \ll 1$.
Then the time derivative of the metric is suppressed compared with
the $y$-derivative of the metric;
\begin{equation}
\left(\frac{\partial_t \alpha}{\partial_y \alpha} \right)^2 
\sim \kappa^2 l \rho \ll 1.
\end{equation}
The bulk metric can be obtained by solving the Einstein equation
(\ref{2-1-4}) and junction conditions (\ref{2-1-5}) perturbatively 
in terms of $\kappa^2 l \rho$. 
The leading order solutions are obtained as \cite{CGRT}
\begin{eqnarray}
\alpha &=& -b(t) \frac{y}{l}+\alpha_0(t),
 \nonumber\\
\beta &=& - b(t) \frac{y}{l} , \nonumber\\
\gamma &=& \log b(t) ,
\label{B}
\end{eqnarray}
where $b(t)$ is the function that describes the time evolution of the 
radion ${\cal R}_0$; 
\begin{equation}
{\cal R}_0 (t) =\int^l_0 dy e^{\gamma(y,t)}=l b(t).
\end{equation}
We assume that the time dependence of $b(t)$ is also weak, i.e. 
$(\dot{b}/l)^2 \sim \kappa^2 l \rho \ll 1$. The behavior of the scale factor 
$\alpha_0$ and the distance between two branes $b(t)$ 
are determined by the next order equations. 
In appendix B, we will obtain the evolution equation for $b(t)$.

\section{Brane displacements and metric perturbations in the bulk}
In the background spacetime, the dynamics of the 4D brane universe 
is independent of the radion ${\cal R}_0(t)$. In order to study 
a cosmological implication of the radion, we shall consider 
linear perturbations about the background spacetime.
As mentioned in the introduction, the radion behaves as the scalar field in 
4D brane universe. Thus we concentrate our attention on 
scalar perturbations. The most general perturbed metric is 
given by
\begin{eqnarray}
ds^2 &=& e^{2 \gamma(y,t)} (1+2 N Q) dy^2 +
e^{2 \beta{(y,t)}} \left(-(1+2 \Phi Q) d t^2 +2 A Q\: dt \: dy \right) 
\nonumber\\
 && +  e^{2 \alpha(y,t)}
\left(\left((1 - 2 D Q) \delta_{ij}+ 2 E Q_{ij}\right) 
dx^i dx^j + 2 B Q_{i} dx^i d t + 2 G Q_{i} dx^i dy \right),
\label{5-2-3}
\end{eqnarray}
where $Q \propto e^{i k^i x_i}$ is the normalized 
harmonics. The vector $Q_i$ and traceless tensor $Q_{ij}$ are constructed 
from $Q$ as 
\begin{equation}
Q_i=- k^{-1} Q_{,i} , \quad Q_{ij}= k^{-2} Q_{,ij}
+\frac{1}{3} \delta_{ij} Q.
\label{5-1-2}
\end{equation}
In the perturbed spacetime, there is a gauge-dependence
in the defeinition of the perturbations. 
Under a first-order coordinate transformation, 
\begin{equation}
x^M \to x^M + \xi^M, \quad \xi^M=(\xi^yQ,\xi^tQ,\xi Q^i),
\label{5-2-1}
\end{equation}
the perturbations transform as 
\begin{eqnarray}
\Phi &=& \hat{\Phi}+\dot{\xi}^t+ \beta' \xi^y + \dot{\beta} \xi^t 
,\nonumber\\
D &=& \hat{D}- \dot{\alpha} \xi^t -\alpha' \xi^y 
-\frac{1}{3} k \xi,\nonumber\\
E &=& \hat{E} - k \xi ,\nonumber\\
B &=&  \hat{B}+\dot{\xi} + k e^{2 (\beta-\alpha)} \xi^t, \nonumber\\
A &=& \hat{A}+e^{2(\gamma-\beta)} \dot{\xi}^y -\xi^{t'} ,\nonumber\\
G &=& \hat{G}-e^{2(\gamma-\alpha)} k \xi^y+\xi' ,\nonumber\\
N &=& \hat{N}+ \xi^{y'}+\dot{\gamma} \xi^t + \gamma' \xi^y.
\label{5-2-4}
\end{eqnarray}  
We should specify the choice of the gauge when we evaluate the 
perturbations.

The most commonly used choice of the gauge to investigate the 
gravitational perturbations in vacuum spacetime is a transverse traceless
(TT) gauge. The perturbed metric in this gauge is given by \cite{Wan2}
\begin{eqnarray}
ds^2 &=&  e^{2 \gamma(y,t)} dy^2 -
e^{2 \beta(y,t)} (1+2 \hat{\Phi} Q) d t^2 \nonumber\\
&& +  e^{2 \alpha(y,t)}
\left(\left((1 - 2 \hat{D} Q) \delta_{ij}+ 2 \hat{E} Q_{ij}\right) 
dx^i dx^j+2 \hat{B} Q_{i} dx^i dt  \right),
\label{5-1-1}
\end{eqnarray}
and the trasverse-traceless conditions are imposed 
on $\hat{\Phi}, \hat{D}, \hat{E}$ and $\hat{B}$ as 
\begin{eqnarray}
\hat{\Phi} &-& 3 \hat{D}= 0, \nonumber\\
 k \hat{B} &+& 2(\dot{\hat{\Phi}}+4 \dot{\alpha} 
\hat{\Phi}) = 0,
\nonumber\\
e^{-2 (\beta-\alpha)}  ( \dot{\hat{B}} 
&+& (-\dot{\beta} +5 \dot{\alpha})
\hat{B} ) - 2 k (\hat{D}+ \frac{2}{3} \hat{E}) = 0 ,\nonumber\\
(\beta' &-& \alpha') \hat{\Phi} = 0.
\label{5-1-4}
\end{eqnarray}
The metric perturbations satisfy a wave equation in the bulk;
 \begin{equation}
-(\hat{E}''+ (-\gamma'+\beta'+3\alpha') \hat{E}')+e^{-2(\beta-\gamma)}
( \ddot{\hat{E}}+(\dot{\gamma}-\dot{\beta}+3 \dot{\alpha}) 
\dot{\hat{E}}) + e^{-2 \alpha} k^2 \hat{E}=0.
\label{5-1-5}
\end{equation}

From the last condition in Eqs. (\ref{5-1-4}), we find that
all components of the perturbations should vanish for $\alpha' \neq \beta'$. 
From the junction conditions Eqs. (\ref{2-1-5}), 
the condition $\alpha'=\beta'$ (which becomes $\alpha_1=\beta_1$ on the brane)
implies $w=w^B=-1$. Thus the TT gauge works only for the cases of 
maximally symmetric 4D branes. We should take other gauges such as a 5D
Longitudinal gauge to solve the bulk perturbations for general cosmological
branes.

However, there is an exception. At large scales, we can 
neglect the terms which are proportional to $k$ compared with the terms
which are the functions of the time derivative or $y$-derivative 
of the metric.
Then we will find the non-zero $\hat{E}$ even for 
$\alpha' \neq \beta'$.
The metric perturbation $\hat{E}$ describes the anisotropy
of the spacetime. Thus the fact that only $\hat{E}$ can be non-zero 
at large scales implies that only anisotropic perturbations are allowed 
to be exist in the bulk for a homogeneous brane. In the following sections, 
we concentrate our attention on the large scale metric perturbations.
Then we take
\begin{equation}
\hat{\Phi}=\hat{D}=\hat{B}=0.
\end{equation}

An important point is that we should allow the 
coordinate location of the branes to acquire a perturbation 
as well as allowing metric perturbations in the bulk. 
When we choose the TT gauge in the bulk, the location of the brane 
is in general displaced \cite{GT}. 
Metric perturbations seen by observer confined to the 
brane are those evaluated at the brane. Thus the 
metric perturbations in TT gauge evaluated at $y=0$
is generally not the observed metric perturbations.
 
Thus we should perform the gauge transformation
to Gaussian normal coordinate system where the normal conditions 
are satisfied $h_{y \mu}=0$ and the branes are held fixed 
at $y=0$ and $y=l$. We refer to this 
coordinate system as brane-GN coordinate \cite{KJ1,KJ2,Wan2}. 
The normal conditions are satisfied by choosing appropriate 
$\xi$ and $\xi^t$;
\begin{eqnarray}
\xi^t &=& \int^y_0 dy e^{2(\gamma-\beta)} \dot{\xi}^y 
 + \epsilon^t(t,x^i), \nonumber\\
\xi &=& k \int^y_0 dy 
 e^{2(\gamma-\alpha)} \xi^y + \epsilon(t,x^i),
\label{5-2-5}
\end{eqnarray}
where $\epsilon^t$ and $\epsilon$ are some functions without $y$ dependence.
These residual gauge transformations 
enable us to impose two additional gauge fixing conditions on the brane. 
Here, we take $\epsilon=\epsilon^t=0$.
After the coordinate transformation, the perturbed metric in
brane-GN coordinate system is given by
\begin{eqnarray}
ds^2 &=& e^{2 \gamma(y,t)} (1+2 N Q) dy^2 -
e^{2 \beta{(y,t)}}(1+2 \Phi Q) d t^2  
\nonumber\\
 && +  e^{2 \alpha(y,t)}
\left(\left((1 - 2 D Q) \delta_{ij}+ 2 E Q_{ij}\right) 
dx^i dx^j + 2 B Q_{i} dx^i d t \right),
\label{5-2-5-1}
\end{eqnarray}
where 
\begin{eqnarray}
\Phi &=&\dot{\xi}^t+ \beta' \xi^y + \dot{\beta} \xi^t 
,\nonumber\\
D &=& - \dot{\alpha} \xi^t -\alpha' \xi^y
-\frac{1}{3}k \xi,\nonumber\\
E &=& \hat{E} - k \xi ,\nonumber\\
B &=& \dot{\xi} + k e^{2 (\beta-\alpha)} \xi^t, \nonumber\\
N &=& \xi^{y'}+\dot{\gamma} \xi^t + \gamma' \xi^y,
\label{QQ}
\end{eqnarray}  
and $\hat{E}$ is the metric perturbations in TT gauge.

The condition that the branes are located at $y=0$ and $y=l$ will be
satisfied by the appropriate choice of $\xi^y$. The function $\xi^y$ evaluated at the 
branes, more precisely, the function
\begin{equation}
\varphi=e^{\gamma_0} \xi^{y}_0,
\label{5-2-7}
\end{equation}
describes the physical displacement of the brane in the TT gauge.  
The point is that the relative displacement of the branes 
$\varphi-\varphi^B$ is the perturbations of the radion at low energies. 
In the spacetime described by the metric (\ref{5-2-5-1}), 
the radion which is the physical distance between two branes 
is given by 
\begin{equation}
{\cal R}(t)=\int^l_0 dy \: e^{\gamma(y,t)} (1+ N(y,t)).
\end{equation}
At low energies, the background spacetime is described by the 
metric (\ref{B}). Then we find $\gamma'=0$ and $\dot{\gamma} \xi^t/\xi^{y'} 
\sim (\kappa^2 \rho l)^2 
\ll 1$ from Eq. (\ref{5-2-5}). Thus the radion ${\cal R}(t)$ is given by
\begin{equation}
{\cal R}(t)= b(t) (l + \xi_0^{yB}-\xi_0^{y})
={\cal R}_0 - (\varphi-\varphi^B).
\end{equation}
Thus the relative displacement of the branes $\varphi-\varphi^B$
can be regarded as the radion at the linear perturbation order.

The induced metric on the brane A is given by
\begin{equation}
ds^2_{brane \:A}=-e^{2 \beta_0}(1+2 \Phi_0 Q) dt^2
+e^{2 \alpha_0}\left((1-2 D_0 Q) \delta_{ij} +2 E_0 Q_{ij} \right) dx^i dx^j.
\label{5-2-6}
\end{equation}
where
\begin{eqnarray}
\Phi_{0} &=& \beta_{1} e^{-\gamma_0} \varphi ,\nonumber\\
D_{0} &=& - \alpha_{1} e^{-\gamma_0} \varphi , \nonumber\\
E_{0} &=& \hat{E}_0.
\label{R-2}
\end{eqnarray}
Metric perturbations in the brane-GN gauge 
coincide with metric perturbations seen by the observer
confined to the brane. 

The metric perturbations should satisfy the junction
conditions at the branes. In the brane-GN coordinate,
the junction conditions can be easily imposed.
We take the perturbed 5D energy-momentum tensor as
\begin{eqnarray}
\delta T^0_0 &=& -\delta \rho \: \delta(y) + \delta \rho^B \: \delta(y-l), 
\nonumber\\
\delta T^i_j &=& \left(\delta p \:\delta(y) - \delta p^B \: \delta(y-l)
\right) \delta^i_j.
\label{5-2-9}
\end{eqnarray}
As in the background spacetime, the junction conditions relate the 
first derivatives of the metric perturbations to the matter perturbations
on the brane \cite{KJ1};
\begin{eqnarray}
\Psi_{1} &=& -\alpha_{1} N_{0} + 
\frac{1}{6} \kappa^2 e^{\gamma_{0}}
\delta \rho, \nonumber\\
\Phi_{1} &=&  \beta_{1} N_{0} + 
\kappa^2 e^{\gamma_{0}} 
\left(\frac{\delta \rho}{3}+
\frac{\delta p}{2} \right), \nonumber\\
E_{1} &=&  0,
\label{5-2-10}
\end{eqnarray}
where $\Psi$ is defined by 
\begin{equation}
\Psi=D-\frac{1}{3}E.
\label{5-2-11}
\end{equation}

It is difficult to find the solutions for metric perturbations
directly in brane-GN coordinate because the bulk evolution equation
is rather complicated in this gauge. Hence, we solve the bulk 
evolution equation in TT gauge and then perform the gauge 
transformation to the brane-GN coordinate. In order to solve the 
bulk perturbation $\hat{E}$ in TT gauge, we should specify the boundary 
conditions for the wave equation (\ref{5-1-5}) for $\hat{E}$. 
Hence, we relate the junction conditions for the metric perturbations 
in brane-GN coordinate Eqs. (\ref{5-2-10}) to the junction conditions 
for the metric perturbations $\hat{E}$ in TT gauge. 
By taking the derivatives of the metric perturbations (\ref{QQ})
with respect to $y$, the first derivatives of the metric perturbations
are written in terms of $\xi^y$ as
\begin{eqnarray}
\Phi_{1} &=& 
e^{2(\gamma_0-\beta_0)} \left(
\ddot{\xi}^{y}_{0} + 
(2 \dot{\gamma}_0 -\dot{\beta}_{0}) \dot{\xi}^{y}_{0} \right)
+ \beta_{1} \xi^{y}_{1}+ \beta_{2} \xi_{0}^{y}, \nonumber\\
\Psi_{1} &=& -\dot{\alpha}_{0}
 e^{2(\gamma_0-\beta_0)}\dot{\xi}^{y}_{0}
- \alpha_{1} \xi^{y}_{1}-\alpha_{2} \xi^{y}_{0}, \nonumber\\
E_{1} &=& \hat{E}_{1} - k^2 e^{-2 (\alpha_{0}-\gamma_{0})} \xi^{y}_{0}.
\label{5-2-13}
\end{eqnarray}
By combining the last equation of Eqs. (\ref{5-2-10}) and Eqs. (\ref{5-2-13}), 
we obtain
\begin{equation}
\hat{E}_{1} = e^{\gamma_{0}-2\alpha_{0}} k^2 \varphi. 
\label{5-2-14}
\end{equation}
This equation gives the boundary condition for 
$\hat{E}(y,t)$ at $y=0$. 
In the similar way, we get the junction condition at the brane B
\begin{equation}
\hat{E}_1^B=e^{\gamma_{0}^B-2 \alpha_{0}^B} k^2 \varphi^B. 
\label{5-2-15}
\end{equation}

The remaining task is to determine the behavior of $\varphi$. 
By combining Eqs. (\ref{5-2-10}) and Eqs. (\ref{5-2-13}) 
and using the equation for $N_0$ (\ref{QQ});
\begin{equation}
N_{0} = \xi_{1}^{y} + \gamma_{1} \xi_{0}^{y},
\label{R-1}
\end{equation}
we can write the matter perturbations
in terms of $\xi^{y}_0$.  We obtain
\begin{eqnarray}
 e^{\gamma_0} \kappa^2 \delta \rho &=& -6 \left( \dot{\alpha}_{0} 
e^{2(\gamma_0-\beta_0)} \dot{\xi}^{y}_{0} 
+(\alpha_{2}-\alpha_{1} \gamma_{1}) \xi^{y}_{0} \right), \nonumber\\
 e^{\gamma_0} \kappa^2 \delta p &=& 2 \left( 
e^{2(\gamma_0-\beta_0)} (
\ddot{\xi}_{0}^{y} + 
(2 \dot{\alpha}_{0} +2 \dot{\gamma}_0-\dot{\beta}_{0})
\dot{\xi}_{0}^{y}) +(2 \alpha_{2}+\beta_2-\beta_1 \gamma_1
-2 \alpha_1 \gamma_1 )\xi_{0}^{y} \right).
\label{5-2-16}
\end{eqnarray}
To simplify the equation, we first rewrite $\alpha_1$, $\alpha_2$, $\beta_1$ 
and $\beta_2$ in terms of $\alpha_0$, $\beta_0$ and $\alpha_1$
using Eq. (\ref{A-1});
\begin{eqnarray}
e^{\gamma_0}\kappa^2 \delta \rho &=& 
-6 e^{2(\gamma_0-\beta_0)} \left(
\dot{\alpha}_0 \dot{\xi}_0^y-(\dot{\alpha}_0^2-\dot{\alpha}_0 
\dot{\gamma}_0) \xi^y_0 \right), \nonumber\\
e^{\gamma_0}\kappa^2 \delta p &=& 2 e^{2(\gamma_0-\beta_0)} \left( 
\ddot{\xi}_0^y+(2 \dot{\alpha}_0+2 \dot{\gamma}_0-\dot{\beta}_0) 
\dot{\xi}^y_0 \right.  \nonumber\\
&+& \left. \left((
-2 \ddot{\alpha_0}+2 \dot{\alpha}_0 \dot{\beta}_0
-3 \dot{\alpha}_0^2 + \ddot{\gamma}_0 + \dot{\gamma}_0^2 -\dot{\gamma}_0 
\dot{\beta}_0+2 \dot{\alpha}_0 \dot{\gamma}_0 )
- e^{2(\gamma_0-\beta_0)}
\frac{(\ddot{\alpha}_0-\dot{\alpha}_0 \dot{\beta}_0)^2}
{\alpha_1^2} 
\right) \xi^y_0 \right). 
\label{5-2-16-1}
\end{eqnarray}
Furthermore, we use the cosmic time $\tau$ on the 
brane defined by $d \tau=e^{\beta_0} dt$ and the physical brane 
displacement $\varphi=e^{\gamma_0} \xi^y_0$.
Then we can write the matter perturbations 
in terms of $\varphi$ as 
\begin{eqnarray}
\kappa^2 \delta \rho &=& -6 \left(\alpha_{0,\tau} \varphi_{,\tau}-
\alpha_{0,\tau}^2 \varphi
\right), \nonumber\\
\kappa^2 \delta p &=& 2 \left(\varphi_{,\tau \tau}+2 \alpha_{0,\tau} 
\varphi_{,\tau}
-\left(2 \alpha_{0,\tau \tau}+3 \alpha_{0,\tau}^2+\frac{\alpha_{0,\tau \tau}^2}
{\alpha_1^2 e^{-2\gamma_0}} \right) \varphi \right). 
\label{5-2-17}
\end{eqnarray}
The matter perturbations
can be written solely by the brane displacement $\varphi$. 
Thus the dynamics of the brane displacement $\varphi$ can be determined 
by specifying the type of the matter perturbations. 
For example, if the matter perturbations are adiabatic, i.e.,
\begin{equation}
\delta p =c_s^2 \delta \rho,
\label{WW}
\end{equation}
we can obtain the evolution equation for $\varphi$.
We should emphasize that the dynamics of the brane displacement 
can be solved independently of the bulk perturbations $\hat{E}$. 
Once the displacement of the 
brane $\varphi$ is derived, the behavior of the trace part
of the metric perturbations $\Phi_0$ and $D_0$ (Eqs.(\ref{R-2})),
and matter perturbations on the brane $\delta \rho$ and $\delta p$ 
(Eqs.(\ref{5-2-17})) can be determined. 
However, one must be careful to consider the fact that the traceless 
part of the metric perturbation $E_0$ cannot be determined. In order to 
know the behavior of $E_0$, we should solve the bulk evolution equation. 

In summary, the problem is to solve the wave equation for
$\hat{E}$ (\ref{5-1-5}) with the boundary conditions 
at the branes (\ref{5-2-14}) and (\ref{5-2-15}). 
The displacement of the brane $\varphi$ 
acts as an source in the boundary conditions. The 
brane displacement $\varphi$ can be determined 
by specifying the type of the matter perturbations
(see Eq. (\ref{5-2-17}) and Eq. (\ref{WW})). 
Once the solution for the brane displacement $\varphi$ is found, 
the boundary conditions for the bulk perturbations $\hat{E}$
are determined. Then solving the wave equations (\ref{5-1-5})
with these boundary conditions, we can find the solution for $\hat{E}_0$.
The observed metric perturbations in brane-GN coordinate are
written in terms of $\hat{E}_0$ and $\varphi$ (\ref{R-2}).
Then the solutions for the metric perturbations on the brane
can be obtained. 

Finally, it should be noted that on the brane A, the anisotropic perturbation 
$E_0$ can be gauged away using the 4D residual gauge transformation 
$\epsilon(t)$ and $\epsilon^t(t)$ in Eqs.(\ref{5-2-5}).  
It is convenient to use the Longitudinal gauge on the brane
in order to compare the result with the conventional 4D general
relativity \cite{4D1,4D2}. In the Longitudinal gauge on the brane, the conditions 
$B^L_0=E^L_0=0$ are imposed. The gauge transformation to the 
Longitudinal gauge from the metric (\ref{5-2-6}) can be carried out by choosing
\begin{eqnarray}
\epsilon &=&  k^{-1} \hat{E}_0, \nonumber\\
\epsilon^t &=&  - k^{-2} e^{2 (\alpha_0-\beta_0)} 
\dot{\hat{E}}_0.
\label{5-2-18}
\end{eqnarray}
The metric perturbations on the brane transform as 
\begin{eqnarray}
\Phi_0^L &=& \Phi_0+\dot{\epsilon}^t+\dot{\beta}_0 \epsilon^t
\nonumber\\
\Psi_0^L &=& D_0- \dot{\alpha}_0 \epsilon^t
\end{eqnarray}
Then, the induced metric in the Longitudinal gauge
on the brane A is given by
\begin{equation}
ds^2_{brane A} = -e^{2 \beta_0} (1+ 2 \Phi^L_0 Q )dt^2 + e^{2 \alpha_0}
(1-2 \Psi^L_0 Q) \delta_{ij} dx^i dx^j, 
\label{5-2-19}
\end{equation}
where
\begin{eqnarray}
\Phi_0^L &=& \beta_1 e^{-\gamma_0} \varphi - k^{-2}
e^{2 \alpha_0} (\hat{E}_{0,\tau \tau} +2 
\alpha_{0, \tau} \hat{E}_{0,\tau}),
\nonumber\\
\Psi_0^L &=& - \alpha_1 e^{-\gamma_0} \varphi 
+ k^{-2} e^{2 \alpha_0} \alpha_{0,\tau} \hat{E}_{0,\tau}
\label{5-2-20}
\end{eqnarray}
The matter perturbations are transformed as 
\begin{eqnarray}
\delta \rho^L &=& \delta \rho -3(1+w) \dot{\alpha}_0 \rho \: \epsilon^t,
\nonumber\\
\delta p^L &=& \delta p -3 c_s^2 (1+w) \dot{\alpha}_0 \rho \:
\epsilon^t.
\end{eqnarray}
Using the evolution equations for background metric Eqs. (\ref{ZZ1})
and (\ref{ZZ2}); 
\begin{equation}
\frac{\alpha_{0,\tau \tau} \alpha_{0 ,\tau}}{\alpha_1 e^{-\gamma_0}}
=(\alpha_1 e^{-\gamma_0})_{,\tau}=
\frac{\kappa^2}{2}(1+w) \alpha_{0,\tau} \rho,
\end{equation}
the matter perturbations in the Longitudinal gauge are given by
\begin{eqnarray}
\kappa^2 \delta \rho^L &=& -6 \left(\alpha_{0,\tau} \varphi_{,\tau}-
\alpha_{0,\tau}^2 \varphi - \frac{\alpha_{0,\tau \tau} \alpha_{0,\tau}}
{\alpha_1 e^{-\gamma_0}} k^{-2} e^{2 \alpha_0} \hat{E}_{0,\tau} 
\right), \nonumber\\
\kappa^2 \delta p^L &=& 2 \left(\varphi_{,\tau \tau}+2 \alpha_{0,\tau} 
\varphi_{,\tau}
-\left(2 \alpha_{0,\tau \tau}+3 \alpha_{0,\tau}^2+\frac{\alpha_{0,\tau \tau}^2}
{\alpha_1^2 e^{-2\gamma_0}} \right) \varphi + 3 c_s^2 
\frac{\alpha_{0,\tau \tau} \alpha_{0,\tau}}{\alpha_1 e^{-\gamma_0}} 
k^{-2} e^{2 \alpha_0} \hat{E}_{0,\tau} \right).
\label{5-2-21}
\end{eqnarray}

\section{Dynamics of brane displacements}
In this section, we investigate the dynamics of the 
displacement of the brane.
For adiabatic perturbations $\delta p=c_s^2 \delta \rho$, 
we can obtain the evolution equation for $\varphi$ 
from (\ref{5-2-17})
\begin{equation}
\varphi_{,\tau \tau}+(2 + 3 c_{s}^2)\alpha_{0,\tau}
\varphi_{,\tau}
-\left(3 \alpha_{0,\tau}^2+2 \alpha_{0,\tau \tau}
+\frac{\alpha_{0,\tau \tau}^2}{\alpha_1^2 e^{-2 \gamma_0}}+3 c_{s}^2
 \alpha_{0,\tau}^2
\right) \varphi =0. 
\label{3-1-19}
\end{equation}
This equation agrees with the evolution equation for radion 
fluctuation derived in Ref \cite{BDL} (see also Appendix B). 
The dynamics of the radion fluctuation can be determined 
independently of the bulk perturbations. 

It is very interesting that this equation has a conserved quantity 
$\zeta_{\ast}$ where $\zeta_{\ast}$ is given by 
\begin{equation}
\zeta_{\ast}=-\alpha_1 e^{-\gamma_0} 
\left[ \varphi -\frac{\alpha_{0,\tau}^2}{\alpha_{0,\tau \tau}}
\left(\frac{1}{\alpha_{0,\tau}} \varphi_{,\tau}-\varphi \right) \right].
\label{3-1-20}
\end{equation}
This can be verified directly by taking the derivative of 
$\zeta_{\ast}$ and using the evolution equation for $\alpha_0$ 
(Eqs. (\ref{ZZ1}) and (\ref{ZZ4})).
Because Eq. (\ref{3-1-20}) gives the first order differential 
equation for $\varphi$, we can integrate the equation to obtain 
the solution for $\varphi$. 
For simplicity we assume $w=c_s^2=const.$
At low energies $\kappa^2 l \rho \ll 1$, the scale factor evolves as
$e^{\alpha_0} \propto \tau^{2/3(1+w)}.$ Then, the equation 
for $\varphi$ is given by
\begin{equation}
\frac{1}{l} \left(
\frac{1+3w}{3(1+w)} \varphi +\tau \varphi_{,\tau} \right)=\zeta_{\ast}.
\end{equation}
The solution for $\varphi$ can be obtained as
\begin{equation}
\varphi= \frac{3(1+w)}{3w+1} l \zeta_{\ast} + f(t) , \quad 
f(t) = f_{\ast} e^{-\frac{3w+1}{2} \alpha_{0}}.
\label{5-3-3}
\end{equation}
At high energies $\kappa^2 l \rho \gg 1$, the scale factor evolves as
$e^{\alpha_0} \propto \tau^{1/3(1+w)}$. The equation for
$\varphi$ is given by
\begin{equation}
\frac{\kappa^2 \rho}{6} \left(\frac{3w+2}{3(1+w)} \varphi + \tau 
\varphi_{, \tau} \right)=\zeta_{\ast}.
\end{equation}
Then we get the solution for $\varphi$ as 
\begin{eqnarray}
\varphi=\frac{6}{\kappa^2 \rho} \frac{3(1+w)}{6w+5} \zeta_{\ast}+f(t) ,
\quad  f(t) = f_{\ast} e^{-(3w+2) \alpha_{0}}
\label{5-3-4}
\end{eqnarray}
where $f_{\ast}$ is again the integration constant. 

We first consider the physical meaning of the 
conserved quantity $\zeta_{\ast}$.
The existence of the conserved quantity indicates the existence
of a conservation's law. To see this fact, it is useful to rewrite
$\zeta_{\ast}$ in terms of the metric perturbations $\Psi_0^L$ and 
the density perturbation $\delta \rho^L$ in the Longitudinal gauge. 
Using Eq. (\ref{5-2-20}) and Eq. (\ref{5-2-21}) we can write $\zeta_{\ast}$
in terms of $\Psi_0^L$ and $\delta \rho^L$;
\begin{equation}
\zeta_{\ast}=\Psi_0^L - \frac{\delta \rho^L}{3(\rho+p)},
\end{equation}
where Eq. (\ref{ZZ3}) is used. 
We should note that bulk perturbation $\hat{E}_0$ in 
$\Psi_0^L$ and $\delta \rho^L$ cancels out in $\zeta_{\ast}$.
Then the conservation of $\zeta_{\ast}$ implies
\begin{equation}
\dot{\zeta_{\ast}}=\dot{\Psi}_0^L - \frac{\dot{\delta \rho^L}}{3 (\rho+p)}
-\dot{\alpha}_0 \frac{\delta \rho^L}{\rho} \frac{1+c_s^2}{1+w}=0,
\end{equation}
where Eq. (\ref{ZZ5}) is used.
This is nothing but the continuity equation for the density perturbation
\begin{equation}
\dot{\delta \rho^L} = 3 (\rho+p) \dot{\Psi}_0^L -3 \dot{\alpha}_0
(\delta \rho^L+ \delta p^L).
\end{equation}
Thus the existence of the conserved quantity in the dynamics of the 
displacement of the brane reflects the fact that the energy momentum-tensor
is conserved in the dynamics of the brane. 
The important point is that the conserved quantity
$\zeta_{\ast}$ is nothing but the conserved curvature perturbation
on hypersurface of uniform energy density.
In the Longitudinal gauge, the curvature perturbation $\zeta$ is 
defined by \cite{Large,4D1,4D2}
\begin{equation}
\zeta=
\Psi_0^L -\frac{\alpha_{0,\tau}^2}{\alpha_{0,\tau \tau}}
\left(\frac{1}{\alpha_{0,\tau}} \Psi_{0,\tau}^L+ \Phi_0^L \right).
\label{5-3-5}
\end{equation}
By substituting Eq. (\ref{5-2-20}) into Eq. (\ref{5-3-5}) 
and using Eq. (\ref{3-1-20}), we can 
find that the curvature perturbation is conserved;
\begin{equation}
\zeta=\zeta_{\ast}.
\end{equation}
In deriving the conservation of the curvature perturbation,
we do not need to solve the bulk perturbations.
This is consistent with the analysis of Ref \cite{4D}. 
In Ref \cite{4D}, it is shown that the conservation of the curvature 
perturbation is derived solely by the conservation of the energy-momentum 
tensor independently of the gravitational field equations. 
\footnote{Here we should note that the entropy perturbation can be induced by the 
bulk gravitational field. In this case, the curvature perturbation is not
conserved and the curvature perturbation cannot be determined unless
the bulk perturbations are solved. 
In the above arguments, the entropy perturbation is implicitly neglected.
Interested readers can refer to the paper \cite{KJ2}.}

Let us examine the coupling between the displacement of the brane $\varphi$
and the matter perturbations on the brane.
From Eqs. (\ref{5-2-17}), we obtain
\begin{equation}
\varphi_{,\tau \tau}+ 3 \alpha_{0,\tau} \varphi_{,\tau}
-\left(2 \alpha_{0,\tau \tau}+4 \alpha_{0,\tau}^2 + 
\frac{\alpha_{0,\tau \tau}^2}{\alpha_1^2 e^{-2 \gamma_0}}  \right)
\varphi = \frac{\kappa^2}{6} \delta T,
\label{5-3-0}
\end{equation}
where $\delta T= -\delta \rho+3 \delta p$. Thus the displacement of the 
brane can be regarded as the scalar field living in the brane 
which couples to the trace of the matter energy-momentum tensor. 
For the de Sitter brane with $\dot{\alpha}_0=H=$const., 
the scalar field has a negative mass squared $-4 H^2$, 
which agrees with the result obtained in Ref \cite{US}. 
By substituting the solution for $\varphi$ 
(\ref{5-3-3}) and (\ref{5-3-4}) into Eq. (\ref{5-3-0}),
we find that the solution for $\varphi$ given by $\zeta_{\ast}$ is 
related to the trace of the matter energy-momentum tensor $\delta T$. 
Thus the solution for $\varphi$ written by $\zeta_{\ast}$
describes the bending of the brane caused by the trace of the matter 
perturbations. It affects the intrinsic curvature $\zeta$
of the brane. 

Next consider the physical meaning of the solution described by $f(t)$. 
The solution $f(t)$ is the homogeneous solution 
for Eq. (\ref{5-3-0}) with $\delta T=0$. Thus it represents
the brane fluctuation which can exist in the absence of 
matter perturbations on the brane. The brane fluctuation does not 
affect the intrinsic curvature of the brane, that is $\zeta$, 
because $f(t)$ is the solution for Eq. (\ref{3-1-20}) with $\zeta=0$.
Thus we find exactly the same phenomena observed in the 
analysis of the thin domain wall fluctuation \cite{VB}
mentioned in the introduction. 
An observer on the brane cannot detect the brane fluctuation
if the observer measures only the intrinsic curvature of the 
brane. However, the important point is that the brane fluctuation
can interact with the bulk perturbation $\hat{E}$ through the 
junction condition for $\hat{E}$ (\ref{5-2-14}). 
Then we find a possibility to detect it. 
In the appendix C, we confirm the fact that the brane fluctuation 
affects the anisotropy of the brane through the interaction with anisotropic 
bulk perturbations. 

\section{Solutions for bulk perturbations}

In the previous section, we found the solution for
the brane displacement $\varphi$. The next step is 
to find the solution for the bulk perturbations
$\hat{E}_0$. As mentioned in section 2, we solve the bulk evolution equation
by the assumption of the nearly static configuration. 
The bulk metric is given by Eq. (\ref{B}). Then
the wave equation for $\hat{E}$ in the bulk (\ref{5-1-5}) is given by
\begin{equation}
\hat{E}''-b(t) \frac{4}{l}  \hat{E}'-b(t)^2 e^{2 b(t)y/l} \left
(\ddot{\hat{E}}+\left(
\frac{\dot{b(t)}}{b(t)}+3 \dot{\alpha}_0-2 \dot{b(t)}\frac{y}{l}
\right) \dot{\hat{E}} \right)=0.
\label{3-3-19}
\end{equation}
The boundary conditions Eqs. (\ref{5-2-14}) and (\ref{5-2-15}) are given by 
\begin{equation}
\hat{E}_1 =  k^{2} b(t) e^{-2 \alpha_0} \varphi(t),\quad 
\hat{E}_1^B= k^{2} b(t) e^{2b(t)} e^{-2 \alpha_0} \varphi^B(t).
\label{Q-3}
\end{equation}

The time dependence of $\hat{E}$ is assumed to be weaker than the 
$y$-dependence of $\hat{E}$; 
\begin{equation}
\left( \frac{\partial_t \hat{E}}{\partial_y \hat{E}} \right)^2 
= {\cal E} \ll 1.
\label{3-3-20}
\end{equation} 
At the leading order, the boundary conditions are assumed to be given by 
$\hat{E}_1=\hat{E}_1^B=0$. Then the solution for $\hat{E}$ is written as 
\begin{equation}
\hat{E}(y,t)=\hat{E}_0(t) + K (t,y),
\label{3-3-21}
\end{equation}
where $K/\hat{E}_0 \sim {\cal E}$.
The wave equation at the next order gives 
\begin{equation}
K'' - b(t) \frac{4}{l} K'=b(t)^2 e^{2 b(t)y/l}
\left(\ddot{\hat{E}}_0+\left( 
\frac{\dot{b(t)}}{b(t)}+3 \dot{\alpha}_0-2 \dot{b(t)}\frac{y}{l} \right) 
\dot{\hat{E}}_0 \right).
\label{3-3-23}
\end{equation}
This equation can be regarded as the ordinal differential equations 
for $K$ with respect to $y$ where the right-hand side acts as a source.
Then the solution for $K(y,t)$ can be written in terms of $\hat{E}_0$;
\begin{equation}
K'(t,y)=m(t)e^{2 b(t)y/l}+n(t) \frac{y}{l} e^{2 b(t) y/l}+p(t)
e^{4 b(t)y/l},
\label{3-3-25}
\end{equation}
where
\begin{eqnarray}
n(t) &=& l b(t) \dot{b(t)} \dot{\hat{E}}_0, \nonumber\\
m(t) &=& -\frac{l}{2} b(t) (\ddot{\hat{E}}_0 + 3 \dot{\alpha}_0 
\dot{\hat{E}}_0),
\label{3-3-26}
\end{eqnarray}
and $p(t)$ is arbitrary. 
On the other hand, the junction conditions for $K(y,t)$ is 
given by (\ref{Q-3}) as
\begin{equation}
K'(0,t)=k^{2} b(t) e^{-2 \alpha_0} \varphi(t),\quad 
K'(l,t)=k^{2} b(t) e^{2b(t)} e^{-2 \alpha_0} \varphi^B(t).
\label{3-3-29}
\end{equation}
These conditions give two constraints on the functions $m(t)$, $n(t)$ and $p(t)$;
\begin{eqnarray}
m(t)+p(t)&=& k^{2} b(t) e^{-2 \alpha_0} \varphi(t), \nonumber\\
m(t)+n(t)+p(t) e^{2 b(t)} &=&  k^{2} b(t) e^{-2 \alpha_0} \varphi^B(t).
\label{3-3-30}
\end{eqnarray}
Then by eliminating $p(t)$ from Eq. (\ref{3-3-30}) and using Eq. (\ref{3-3-26})
we obtain the evolution equation for $\hat{E}_0$ as
\begin{equation}
\ddot{\hat{E}}_0+ \left( 3 \dot{\alpha}_0 + \dot{b}(t) \frac{e^{-b(t)}}
{\sinh b(t)}\right) \dot{\hat{E}}_0 = -k^{2} \frac{e^{b(t)}}
{\sinh b(t)} e^{-2 \alpha_0}  l^{-1}
\left(\varphi(t)-e^{-2 b(t)} \varphi^B(t) \right).
\label{3-3-31}
\end{equation}
The equation (\ref{3-3-31}) can be integrated to give the 
solution for $\dot{\hat{E}}_0$;
\begin{equation}
\dot{\hat{E}}_0= - k^{2} e^{-3 \alpha_0} 
\left(\frac{e^{b(t)}}{\sinh b(t)} \right) \int 
e^{\alpha_0} l^{-1}
\left(\varphi(t)-e^{-2b(t)} \varphi^B(t) \right) \: dt.
\label{3-3-34}
\end{equation}
This can be verified using the equation
\begin{equation}
\left(\frac{e^{b(t)}}{\sinh b(t)} \right)^{\cdot}
= -\dot{b}(t) \frac{e^{-b(t)}}{\sinh b(t)} 
\left(\frac{e^{b(t)}}{\sinh b(t)} \right).
\label{3-3-33}
\end{equation}


\section{Metric perturbations on the brane}
\subsection{Solutions for metric perturbations on the brane}
Once we obtain the solution 
for $\varphi$ (\ref{5-3-3}) and $\hat{E}_0$ (\ref{3-3-34}), 
the behavior of the metric perturbations 
and density perturbation in the Longitudinal gauge 
can be derived from Eqs. (\ref{5-2-20}) and (\ref{5-2-21}).
At the leading order of the low energy approximation,
we get
\begin{eqnarray}
\Psi_0^L &=& \frac{1}{l} \varphi+ k^{-2}\dot{\alpha}_0 
e^{2 \alpha_0} \dot{\hat{E}}_0, \nonumber\\
\Phi_0^L &=& -\frac{1}{l} \varphi- k^{-2} e^{2 \alpha_0}\left(
\ddot{\hat{E}}_0 + 2 \dot{\alpha}_0 \dot{\hat{E}}_0 \right), 
\nonumber\\
\kappa^2  \delta \rho^L &=&-6
\left(\dot{\alpha}_0 \dot{\varphi} - \dot{\alpha}_0^2 \varphi 
+l \dot{\alpha}_0 \ddot{\alpha}_0 e^{2 \alpha_0} k^{-2}
\dot{\hat{E}}_0 \right).
\label{6-1-5}
\end{eqnarray}
In the following section, we explicitly derive the solution for
metric perturbations and consider the observational implications.

\subsection{One brane model}
First let us consider the one brane model. 
By taking $b \to \infty$,
the solution for $\hat{E}_0$ (\ref{3-3-34}) is given by 
\begin{equation}
k^{-2} \dot{\hat{E}}_0=-l \frac{18(1+w)^2}{(3w+5)(3w+1)} \zeta_{\ast}
\left(\frac{t}{l} \right)^{\frac{3w-1}{3(1+w)}}
- \frac{3(1+w)}{2} f_{\ast} 
\left(\frac{t}{l} \right)^{-\frac{2}{3(1+w)}}
+F l \left(\frac{t}{l} \right)^{-\frac{2}{1+w}},
\label{6-2-1}
\end{equation}
where $F$ is the integration constant and we used the 
solution for $\varphi$ (Eq. (\ref{5-3-3})) and for the scale factor;
\begin{equation}
e^{\alpha_0}=\left(\frac{t}{l} \right)^{\frac{2}{3(1+w)}}.
\label{6-2-2}
\end{equation} 
Now substituting the solutions for $\varphi$ and $\hat{E}_0$
into Eq. (\ref{6-1-5}), we can determine
the behavior of the perturbations on the brane
\begin{eqnarray}
\Psi_0^L &=& \Phi_0^L= \frac{3(1+w)}{3w+5} \zeta_{\ast} +\tilde{F} 
\left( \frac{t}{l}
\right)^{-\frac{3+5w}{3(1+w)}}, \nonumber\\
\frac{\delta \rho^L}{\rho} &=& -2 \Psi_0,
\label{6-2-3}
\end{eqnarray}
where $\tilde{F}=2F /3(1+w)$. This 
agrees with the result obtained in the conventional 4D 
Einstein theory. It should be emphasized that the brane fluctuation 
$f(t)$ does not appear in the metric perturbations on the brane. 

\subsection{Two brane model with time independent distance}
Next let us consider the two branes model where the distance between  
two branes is time independent, i.e. $b(t)=b_{\ast}=const$. 
In Appendix B, we show that
the fine-tuning between the energy density of the matter
on the brane A and the energy density of the matter on the brane B is needed 
to make the distance constant. Then the brane A and the brane B should expand in 
the same way $e^{\alpha_0} \propto e^{\alpha_0^B}$ for this condition. 
  
The solution for $\hat{E}_0$ is obtained as
\begin{eqnarray}
k^{-2} \dot{\hat{E}}_0 &=& -l \frac{18(1+w)^2}{(3w+5)(3w+1)} \left( 
\frac{e^{b_{\ast}}}{2 \sinh b_{\ast}} \right)
(\zeta_{\ast}-e^{-2 b_{\ast}} \zeta_{\ast}^B) 
\left( \frac{t}{l} \right)^{\frac{3w-1}{3(1+w)}} \nonumber\\
&-&
 \frac{3(1+w)}{2} \left(
\frac{e^{b_{\ast}}}{2 \sinh b_{\ast}} \right)
(f_{\ast}-e^{-2 b_{\ast}}f_{\ast}^B)  
\left( \frac{t}{l} \right)^{
-\frac{2}{3(1+w)}} 
+ F l \left(\frac{t}{l} 
\right)^{-\frac{2}{1+w}}.
\label{6-3-1}
\end{eqnarray}
Then substituting the solution for $\hat{E}_0$ into Eq. (\ref{6-1-5}), 
the metric perturbations are given by
\begin{eqnarray}
\Psi_0^L &=& \frac{3(1+w)}{3w+1} \zeta_{\ast} \left(
\frac{3w+1}{3w+5}
- {\cal N}(b_{\ast}) \frac{4}{3w+5} \right)
+ {\cal N}(b_{\ast}) \frac{12(w+1)}{(3w+1)(3w+5)} \zeta_{\ast}^B \nonumber\\
&& - {\cal N}(b_{\ast}) l^{-1}(f(t)-f^B(t))
 +\tilde{F} \left( \frac{t}{l}\right)^{-\frac{3+5w}{3(1+w)}}, \nonumber\\
\Phi_0^L &=& \frac{3(1+w)}{3w+1} \zeta_{\ast} \left(
\frac{3w+1}{3w+5}
+ {\cal N}(b_{\ast}) \frac{6(1+w)}{3w+5} \right)
-{\cal N}(b_{\ast}) \frac{18(w+1)^2}{(3w+1)(3w+5)} 
\zeta_{\ast}^B \nonumber\\
&& + {\cal N}(b_{\ast}) l^{-1}(f(t)-f^B(t))
 +\tilde{F} \left( \frac{t}{l}\right)^{-\frac{3+5w}{3(1+w)}},
\label{6-3-2}
\end{eqnarray}
where
\begin{equation}
{\cal N}(b_{\ast})=\frac{e^{-b_{\ast}}}{2 \sinh b_{\ast}}.
\end{equation}

There are three types of the modifications in the two branes model.
First, the relation between the metric perturbations
and the conserved curvature perturbation is modified from the
one brane model.
Secondly, the "shadow matter" appears, which can be seen
from the second term in Eq. (\ref{6-3-2}). The curvature perturbation
on the brane B is projected onto the brane A. 
These phenomena are already known 
from the analysis of the weak gravitational field in the model with two
Minkowski branes \cite{GT}. 
It has been shown that the linearized gravity on the brane
becomes Brans-Dicke (BD) type gravity due to these modifications.  
The BD parameter $w_{BD}$ is related to the distance between two branes 
$b_{\ast}$ as
\begin{equation}
w_{BD}=\frac{3}{2}(e^{2 b_{\ast}}-1).
\end{equation}
From the observations, we need $w_{BD} > 3000$ at present, 
which implies $b_{\ast}>4$. 
The modifications are suppressed by the factor ${\cal N}(b_{\ast})$.
From the constraint $b_{\ast} > 4$, this suppression factor is 
estimated as 
\begin{equation}
{\cal N}(b_{\ast})=\frac{e^{-b_{\ast}}}{2 \sinh b_{\ast}}< 3 \times 10^{-4}.
\end{equation}
Finally, the radion caused by the relative difference between the 
brane fluctuation of the brane A and the brane B, $f(t)-f^B(t)$,
contributes to the metric perturbations. 
This effect is also suppressed by the factor ${\cal N}(b_\ast)$.
 
Let us consider the large scale temperature fluctuations in CMB caused by 
the ordinary Sachs-Wolfe effect. The temperature anisotropy
is given by \cite{Large}
\begin{equation}
\frac{\Delta T}{T}=-\zeta+ \Psi_0^L + \Phi_0^L.
\end{equation}
Then the modifications of the temperature anisotropy due to the existence
of the hidden brane are given by
\begin{equation}
\frac{\Delta T}{T}(b_{\ast}) = {\cal N}(b_{\ast}) \left(
\frac{6(w+1)}{3w+5} \right)(\zeta_{\ast}- \zeta_{\ast}^B).
\end{equation}
It should be mentioned that the radion caused by brane fluctuations 
does not affect the temperature anisotropy. 

\subsection{Two branes with time dependent distance}
Now, we consider the general situation where the 
distance between two branes varies with time.
We take $\varphi^B=0$ and neglect the 
decaying mode of the solution for simplicity.
Then the solution for $\hat{E}_0$ is given by
\begin{equation}
k^{-2} \dot{\hat{E}}_0= -l\left(\frac{e^{b(t)}}{2\sinh(b(t))} \right)
\left( \frac{18(1+w)^2}{(3w+5)(3w+1)} \zeta_{\ast}
\left(\frac{t}{l} \right)^{\frac{3w-1}{3(1+w)}} 
+ \frac{3(1+w)}{2} l^{-1} f_{\ast}
\left(\frac{t}{l} \right)^{-\frac{2}{3(1+w)}} \right).
\end{equation}
The metric perturbations are obtained as 
\begin{eqnarray}
\Psi_0^L &=& \frac{3(1+w)}{3w+1} \left( 
\frac{3w+1}{3w+5} -  {\cal N}(b(t))
\frac{4}{3w+5} \right) \zeta_{\ast} - {\cal N}(b(t)) l^{-1} 
f(t) ,
\nonumber\\
\Phi_0^L &=& \left[ \frac{3(1+w)}{3w+1} 
\left( \frac{3w+1}{3w+5} +{\cal N}(b(t)) 
\frac{6(1+w)}{3w+5} \right)- \frac{\dot{{\cal N}}(b(t))}{\dot{\alpha}_0}  
\frac{12 (1+w)}{(3w+1)(3w+5)} \right]  \zeta_{\ast} \nonumber\\
&& + \left[ {\cal N}(b(t))
+ \frac{\dot{{\cal N}}(b(t))}{\dot{\alpha}_0} \right] l^{-1} f(t),  
\end{eqnarray}
where
\begin{equation}
{\cal N}(b(t))=\frac{e^{-b(t)}}{2 \sinh (b(t))}.
\end{equation}
A new effect is induced if the distance between two branes
varies with time. 
There are terms proportional to the time derivative 
of $b(t)$.
The time variation of $b(t)$ makes the effective 4D 
Newton constant $G_4$ vary with time.
The effective 4D Newton constant $G_4$ on the positive tension brane A is
determined by the 5D Newton constant by (see Ref.\cite{GT})
\begin{equation}
G_4 =\frac{G_5}{l} \frac{e^{b(t)}}{2 \sinh b(t)}.
\end{equation}
The time evolution of the 4D Newton constant is then given by
\begin{equation}
\frac{\dot{G_4}}{G_4}= \dot{\cal{N}}(b(t)).
\end{equation}
The time variation of the Newton constant is constrained 
by several experiments. For example, the successful Big-Bang
nucleosynthesis requires $\dot{G_4}/G_4 \sim 10^{-11} \mbox{yr}^{-1}$. 
Thus the terms proportional to the time variation of $b(t)$ are
suppressed by the factor
\begin{equation}
\frac{\dot{{\cal N}}(b(t))}{\dot{\alpha}_0} \sim
\frac{\dot{G_4}}{G_4 H_0} \frac{H_0}{H(t)} \sim 10^{-1} 
\frac{H_0}{H(t)},
\end{equation}
where $H_0^{-1} \sim 10^{10}$ yr is the present day lifetime of the 
universe and $H(t)=\dot{\alpha_0}(t)$ is the Hubble scale at $t$.
The modifications of the temperature anisotropy $\Delta T/T$ 
due to the time variation of $b(t)$ are given by
\begin{equation}
\frac{\Delta T}{T}(\dot{b(t)}) 
= \frac{\dot{{\cal N}}(b(t))}{\dot{\alpha_0}}
\left(l^{-1} f(t) -\frac{12(1+w)}{(3w+1)(3w+5)} \zeta_{\ast} \right).
\label{Q-5}
\end{equation}
We should notice that the brane fluctuation contributes to the temperature
anisotropy. 
However, the factor $\dot{\cal{N}}(b(t))/\dot{\alpha}_0$ at the 
decoupling is significantly suppressed like
\begin{equation}
\left(\frac{\dot{{\cal N}}(b(t))}{\dot{\alpha}_0}\right)_{dec} \sim 10^{-6}.
\end{equation}
where we used $H_0/H(t_{dec}) \sim 10^{-5}$. 
In addition, the brane fluctuation $f(t)$ decreases with time in a dust 
dominated universe. 
Then it seems to be impossible to detect the fluctuation of our brane 
in the CMB anisotropy.

If we take into account the displacement of the brane B, the situation 
becomes more complicated. The contribution from the displacement of the brane
B in the CMB temperature anisotropy on our brane is given by
\begin{equation}
\frac{\Delta T}{T}(\varphi^B) =2
\left(2 \dot{\alpha}_0 e^{2 b(t)} {\cal N}(b(t)) + \dot{{\cal N}}(b(t)) \right) e^{-\alpha_0}
\int dt e^{\alpha_0} e^{-2 b(t)} l^{-1} \varphi^B + 2 {\cal N}(b(t)) l^{-1} 
\varphi^B.
\end{equation}
If we allow that the distance between two branes varies with time, 
the history of the universe on the brane B can be different from 
the history of our universe. Thus unless we specify the history 
of the hidden brane B, this contribution cannot be determined. 

\section{Summary and Discussions}
In this paper, we investigated a cosmological implication of the radion. 
We considered the $S^1/Z_2$ compactified 5D AdS spacetime. 
Two positive and negative tension branes are located at the orbifold 
fixed points. We are assumed to be living on the positive tension brane.
At the linear perturbation order, the radion is the relative displacement 
of the brane.

There are two different kinds of the displacements 
of the brane. One describes a fluctuation of the 
brane which does not couple to matter perturbations on the brane.
The other describes a bend of the
brane which couples to matter perturbations on the brane.

The bend of the brane is described by the conserved quantity 
which appears in the dynamics the displacement of the brane.
It was shown that the conserved quantity is related to
the conservation of the energy-momentum tensor on the brane. The
point is that this conserved quantity is nothing but the conserved
curvature perturbation on uniform density hypersurface. Thus the 
conservation of the curvature perturbation can be derived solely
by the evolution equation for the brane displacement, which is independent 
of the bulk perturbations. Thus this confirms the recent claim that the
curvature perturbation at large scales is conserved in any gravitational theory 
if the energy-momentum tensor is conserved. 
We also found a solution for the brane fluctuation. 
For a cosmological brane, the homogeneous brane fluctuation $f(t)$ evolves 
as $f(t) \propto e^{-(3w+1) \alpha_0/2}$ at low energies 
where $e^{\alpha_0(t)}$ is the scale factor of the brane and $w$ is 
the constant barotropic index of the matter. 
It was shown that the brane fluctuation does not affect the curvature 
perturbation.

As well as allowing the displacement of the brane, 
we should allow the bulk to acquire perturbations. We found that
at large scales, only anisotropic perturbations are allowed 
to be exist. Then the homogeneous brane displacements interact
with anisotropic perturbations in the bulk. Assuming that the
system is nearly static, we found the solution for 
the bulk anisotropic perturbations induced by brane displacements.  

The metric perturbations on the brane in the Longitudinal gauge are written by 
the displacement of the brane and the anisotropic bulk perturbation induced 
by the displacements. 
If the distance between two branes is time independent, three types of the 
modifications are found in the solution for the  
metric perturbations on our brane compared with the conventional 
4D Einstein theory. First, the relation between the 
metric perturbation and the conserved curvature perturbation is modified.
Secondly, the curvature perturbation on the hidden brane appears in the 
solution for the metric perturbations. Finally, the radion 
caused by the relative difference between the fluctuation of 
our brane and the fluctuation of the hidden brane contributes
to the metric perturbations. It is well known that 
the linearized gravity becomes Brans-Dicke gravity due to these
modifications. The Brans-Dicke parameter is determined by the distance
between two branes. The observational
constraints on the Brans-Dicke parameter gives the constraint on the distance.
Then all these modifications have the same suppression 
factor of the order $10^{-4}$. An interesting finding is that the 
radion caused by brane fluctuations does not affect the CMB anisotropy 
if the distance is time independent.

If the distance between two branes has time dependence, 
modifications which are proportional to the time derivative of the distance
will appear. In this case, the brane fluctuation can affect the
CMB anisotropy. However, the time variation of the distance between 
two branes makes the effective 4D Newton constant vary with time. The time
variation of the 4D Newton constant is constrained by the
observations. If we consider only the displacement of our brane, 
the modification induced by the time variation of the 
distance is suppressed at the decoupling by the 
factor of the order $10^{-6}$. 
Because the brane fluctuation $f(t)$ is the decreasing function in a dust 
dominated universe, it is hard to detect the fluctuation of our brane. 
On the other hand, if we consider the 
displacements of the hidden brane, the situation becomes more complicated. 
The displacements of the hidden brane appears in the CMB anisotropy on our brane
in a non-trivial manner. Unlike the displacements of our brane, 
the displacements of the hidden brane are not directly constrained 
by the observation because the history of the universe on the hidden brane 
needs not be the same as the history of our universe. The amplitude of the
curvature perturbation and the fluctuation of the hidden brane can be of the 
order one. Thus there is 
a possibility to detect the existence of the hidden brane in the 
CMB anisotropy. It will be very interesting to perform detailed calculations
with some specific models. Then it may be give the answer for the question
whether our universe can be realized in two branes models.

\appendix
\section{Equations for background metric}
From 5D Einstein equations, the equations for the metric $\alpha$, $\beta$ and 
$\gamma$ in the bulk are given by
\begin{eqnarray}
-\frac{2 e^{2 \gamma}}{l^2} &=& 
e^{-2(\beta-\gamma)} (\dot{\alpha}^2+ \dot{\alpha} \dot{\gamma})
+(-\alpha''-2 \alpha'^2+\alpha' \gamma'), \label{Q-1} \\
\frac{2 e^{2 \gamma}}{l^2} &=&
e^{-2(\beta-\gamma)}(-\ddot{\alpha}-2 \dot{\alpha}^2+\dot{\alpha} \dot{\beta})
+\alpha'^2+ \alpha'  \beta' , \label{C} \\
\frac{6 e^{2 \gamma}}{l^2} &=& 
e^{-2(\beta-\gamma)} 
(-2 \ddot{\alpha}-\ddot{\gamma}-3 \dot{\alpha}^2+\dot{\beta} \dot{\gamma} -
\dot{\gamma}^2+2 \dot{\alpha} \dot{\beta}-2 \dot{\alpha} \dot{\gamma}) 
\nonumber\\
&&+ (2 \alpha''+\beta''+3 \alpha'^2 +\beta'^2-\beta'\gamma'-2 \alpha'
 \gamma'+2 \alpha' \beta'), \label{Q-2}\\
0 &=& \beta' \dot{\alpha}- \dot{\alpha}'-\dot{\alpha} \alpha'+\alpha' 
\dot{\gamma}. \label{D}
\end{eqnarray}
The first and the second equation can be integrated with the help of
the last equation as 
\begin{equation}
e^{-2 \gamma} \alpha'^2-e^{-2 \beta} \dot{\alpha}^2 =\frac{1}{l^2}.
\label{A-2}
\end{equation}

First we write $\alpha_1$ and $\beta_1$ by $\alpha_0, \beta_0$
and $\gamma_0$. Evaluating Eq. (\ref{A-2}) on the brane, 
$\alpha_1^2$ can be written as
\begin{equation}
\alpha_1^2 
= \frac{e^{2 \gamma_0}}{l^2} + e^{-2(\beta_0-\gamma_0)} \dot{\alpha}_0^2.
\label{A-3}
\end{equation}
From Eqs. (\ref{C}) and (\ref{A-2}), $\beta_1$ 
is related to $\alpha_1$;
\begin{equation}
\beta_1= \alpha_1 \left(1+ e^{-2(\beta_0-\gamma_0)} \frac{\ddot{\alpha}_0
-\dot{\alpha}_0 \dot{\beta}_0}{\alpha_1^2} \right).
\label{A-4}
\end{equation}
Next, we write $\alpha_2$ and $\beta_2$ by $\alpha_0,
\beta_0,\gamma_0$ and $\alpha_1$. 
The $y^0$-th order power series expansion of 
Eq. (\ref{Q-1}) and Eq. (\ref{Q-2}) give
\begin{eqnarray}
\alpha_2-\alpha_1 \gamma_1 &=& e^{-2(\beta_0-\gamma_0)} (\dot{\alpha}_0^2+
\dot{\alpha}_0 \dot{\gamma}_0) -2 \alpha_1^2 + \frac{2 e^{2 \gamma_0}}{l^2}
\nonumber\\
2 \alpha_2+\beta_2-\beta_1 \gamma_1-2 \alpha_1 \gamma_1 &=& 
-3 \alpha_1^2-\beta_1^2-2 \alpha_1 \beta_1, \nonumber\\
&&+ e^{-2(\beta_0-\gamma_0)} (2 \ddot{\alpha}_0 +\ddot{\gamma}_0
+3 \dot{\alpha}_0^2-\dot{\beta}_0 \dot{\gamma}_0+\dot{\gamma}_0^2
-2 \dot{\alpha}_0 \dot{\beta}_0+2 \dot{\alpha}_0 \dot{\gamma}_0) \nonumber\\
&& +\frac{6 e^{2 \gamma_0}}{l^2}. 
\end{eqnarray}
Then substituting the equtions for $\alpha_1$ (Eq. (\ref{A-3})) and 
$\beta_1$ (Eq. (\ref{A-4})), we get 
\begin{eqnarray}
\alpha_2-\alpha_1 \gamma_1 &=& e^{2(\gamma_0-\beta_0)}(-\dot{\alpha}_0^2+
\dot{\alpha}_0 \dot{\gamma}_0), \nonumber\\
2 \alpha_2+\beta_2-\beta_1 \gamma_1-2 \alpha_1 \gamma_1 &=&
e^{2(\gamma_0-\beta_0)} (-2 \ddot{\alpha_0}+2 \dot{\alpha}_0 \dot{\beta}_0
-3 \dot{\alpha}_0^2 + \ddot{\gamma}_0 + \dot{\gamma}_0^2 -\dot{\gamma}_0 
\dot{\beta}_0+2 \dot{\alpha}_0 \dot{\gamma}_0 ) \nonumber\\
&-& e^{4(\gamma_0-\beta_0)}
\frac{(\ddot{\alpha}_0-\dot{\alpha}_0 \dot{\beta}_0)^2}
{\alpha_1^2}.  
\label{A-1}
\end{eqnarray}

Let us derive some useful equations for $\alpha$.
Eq. (\ref{D}) gives
\begin{equation}
\dot{\alpha}_1-\alpha_1 \dot{\gamma}_0 =
\dot{\alpha}_0 (\beta_1 -\alpha_1) 
= e^{-2(\beta_0-\gamma_0)} \frac{(\ddot{\alpha}_0-\dot{\alpha}_0 \dot{\beta}_0)
\dot{\alpha}_0}{\alpha_1},
\label{A-0}
\end{equation}
where we used Eq. (\ref{A-4}).
Using the cosmic time $\tau$ this equation can be written as 
\begin{equation}
(\alpha_1 e^{-\gamma_0})_{,\tau}=
\frac{\alpha_{0,\tau \tau} \alpha_{0 ,\tau}}{\alpha_1 e^{-\gamma_0}}.
\label{ZZ1}
\end{equation}
On the other hand, from the junction condition and the 
conservation's law for the energy density, we obtain 
\begin{equation}
(\alpha_1 e^{-\gamma_0})_{,\tau}= -\frac{\kappa^2}{6} \rho_{, \tau}
=\frac{\kappa^2}{2}(1+w)\alpha_{0,\tau} \rho.
\label{ZZ2}
\end{equation}
Then the second derivative of $\alpha_0$ with respect to $\tau$
is given by
\begin{equation}
\alpha_{0,\tau \tau}= \frac{\kappa^2}{2}(\alpha_1 e^{-\gamma_0})
(1+w) \rho.
\label{ZZ3}
\end{equation}
Using the equation
\begin{equation}
w_{,\tau}=3(1+w)(w-c_s^2) \alpha_{0,\tau},
\label{ZZ5}
\end{equation}
we can also obtain the third derivative of $\alpha_0$ with respect to $\tau$
as 
\begin{equation}
\alpha_{0,\tau \tau \tau} = \left( 
\frac{\alpha_{0,\tau \tau}}{\alpha_1 e^{-\gamma_0}}
\right)^2 \alpha_{0,\tau} -3(1+c_s^2) \alpha_{0,\tau} \alpha_{0,\tau \tau}.
\label{ZZ4}
\end{equation}

\section{The dynamics of the distance between two branes}
In this appendix, we derive the evolution equation for 
$b(t)$. The 5D metric is given by 
\begin{eqnarray}
\alpha &=& -b(t) \frac{y}{l}+\alpha_0(t)+\bar{\alpha}(y,t),
 \nonumber\\
\beta &=& - b(t) \frac{y}{l}+\bar{\beta}(y,t) , \nonumber\\
\gamma &=& \log b(t) + \bar{\gamma}(y,t),
\label{2-1-10}
\end{eqnarray}
where $\bar{\alpha}$, $\bar{\beta}$ and $\bar{\gamma}$ are the functions 
of the order ${\cal O}(\kappa^2 l \rho)$. 
The junction conditions of the order ${\cal O}(\kappa^2 \rho l)$ give 
the junction conditions for $\bar{\alpha}(y,t)$ and $\bar{\beta}(y,t)$ as 
\begin{eqnarray}
\bar{\alpha}_{1}(t) &=& 
-\frac{1}{l} b(t) \left(\bar{\gamma}_0+ \frac{\kappa^2 l \rho}{6} \right), \nonumber\\
\bar{\beta}_{1}(t) &=& 
-\frac{1}{l} b(t) \left(\bar{\gamma}_0- \left(\frac{\kappa^2 l \rho}{3}
 +\frac{\kappa^2 l p}{2} \right) \right). 
\label{2-1-11}
\end{eqnarray}
We need not solve the next order solution $\bar{\alpha}, \bar{\beta}$ and 
$\bar{\gamma}$ in order to derive the evolution equation for 
$\alpha_0(t)$ and $b(t)$.
Einstein equations (\ref{2-1-4}) at the next order in $\kappa^2 l \rho$ 
give 
\begin{eqnarray}
(\dot{\alpha}_0-\dot{b}(t) y/l)^2 e^{2 b(t) y/l} &=& 
- \left( \frac{2}{l^2} \bar{\gamma}+ \frac{2}{b(t)l} \bar{\alpha}' \right),
\nonumber\\
\dot{\bar{\alpha}}' &=& \frac{\dot{b}(t)}{b(t)} \bar{\alpha}'- \frac{b(t)}{l}
\dot{\bar{\gamma}} +(\dot{\alpha}_0(t)-\dot{b}(t) y/l)(\bar{\beta}'-\bar{\alpha}').
\label{2-1-12}
\end{eqnarray}
Evaluating Eqs. (\ref{2-1-12}) on the brane A and substituting the 
junction conditions (\ref{2-1-11}), we get 
\begin{eqnarray}
\dot{\alpha}_0^{2} &=& \frac{\kappa^2 \rho}{3 l}, \nonumber\\
\dot{\rho} &+& 3(1+w) \dot{\alpha}_0 \rho = 0.
\label{2-1-13}
\end{eqnarray}
In the same way, Eqs. (\ref{2-1-13}) evaluated at the brane B give
\begin{eqnarray}
\left(\dot{\alpha}_0 - \dot{b} \right)^2 e^{2 b(t)} &=& -
\frac{\kappa^2 \rho^B}{3 l}, \nonumber\\
\dot{\rho}^B &+& 3(1+w^B) \left(\dot{\alpha}_0 -\dot{b} \right)
\rho^B = 0. 
\label{2-1-14}
\end{eqnarray}
From Eqs. (\ref{2-1-13}) and (\ref{2-1-14}), 
we can determine $b(t)$ and $\alpha_0(t)$. 

The evolution equation for the displacement of the brane can be also 
derived from the evolution equation for $b(t)$ \cite{BDL}. 
From Eqs. (\ref{2-1-13}) and (\ref{2-1-14}),
we find that if the matter on the branes 
satisfies the conditions 
\begin{equation}
w=w^B,\quad \rho=- \rho^B e^{-2 b_{\ast}},
\label{2-2-1}
\end{equation}
the distance between two branes becomes constant; $b(t)=b_{\ast}=const$. 
Let us consider the fluctuation of the energy density of the
matter on the brane A;
\begin{equation}
\rho =-\rho^B  e^{-2 b_{\ast}} + \delta \rho,
\label{2-2-2}
\end{equation}
where we assume $\delta \rho/\rho \ll 1$.  
Then the brane A is displaced. We define the displacement of the brane 
$\varphi$ as 
\begin{equation}
b(t)=b_{\ast}-\varphi(t)/l,
\label{2-2-3}
\end{equation}
where $\varphi/l b_{\ast} \ll 1$. 
Let us obtain the evolution equation for
$\varphi$.  From Eq. (\ref{2-1-14}), the equation for 
$\varphi$ is given by
\begin{equation}
\left(\dot{\varphi}/l+\dot{\alpha}_0 \right)^2 e^{-2 \varphi}
=-\frac{\kappa^2 \rho^B e^{-2 b_{\ast}}}{3 l}.
\label{2-2-4}
\end{equation}
By linearizing the equation with respect to $\varphi$, we find
\begin{equation}
\dot{\alpha}_0 \dot{\varphi}-\dot{\alpha}_0^2 \varphi=-\frac{\kappa^2}{6}
(\rho^B e^{-2 b_{\ast}}+\rho)=-\frac{\kappa^2}{6} \delta \rho,
\label{2-2-5}
\end{equation}
where we used Eqs. (\ref{2-1-13}) and (\ref{2-2-2}).
Taking the time derivative of Eq. (\ref{2-2-5}) and using the 
equations 
\begin{eqnarray}
\ddot{\alpha}_0 \dot{\varphi} &=& -\frac{\kappa^2 \rho}{2}
(1+w)
\dot{\varphi} \sim \frac{\kappa^2 \rho^B e^{-2 b_{\ast}}}{2}
(1+w^B) \dot{\varphi}, \nonumber\\
\dot{\delta \rho} &=& -3 \dot{\alpha}_0 (\delta \rho+ \delta p)
-3 \dot{\varphi} (1+w^B) \rho^B  e^{-2 b_{\ast}},
\label{2-2-6}
\end{eqnarray}
which can be derived from $\delta \rho= \rho+\rho^B e^{-2 b_{\ast}}$ 
and Eq. (\ref{2-1-13}) and (\ref{2-1-14}), we get
\begin{equation}
\ddot{\varphi} -\dot{\alpha}_0 \dot{\varphi} -2 
\ddot{\alpha}_0 \varphi = \frac{\kappa^2}{2} (\delta \rho
+\delta p),
\label{2-2-7}
\end{equation}
where we defined $\delta p=w \delta \rho$. 
Then we can write matter perturbations in terms of $\varphi(t)$ 
\begin{eqnarray}
\kappa^2 \delta \rho &=& -6(\dot{\alpha}_0 \dot{\varphi}
 -\dot{\alpha}_0^{2} \varphi ), \nonumber\\
\kappa^2 \delta p &=& 2(\ddot{\varphi}+2 \dot{\alpha}_0 \dot{\varphi} 
-(2\ddot{\alpha}_0+ 3 \dot{\alpha}_0^{2}) \varphi ).
\label{2-2-8}
\end{eqnarray}
Using $\delta p=w \delta \rho$, we can 
obtain the evolution equation for $\varphi$ as 
\begin{equation}
\ddot{\varphi}+(2+3 w) \dot{\alpha}_0 \dot{\varphi} -
(3 \dot{\alpha}_0^{2}+2 \ddot{\alpha}_0 ) \varphi=0.
\label{2-2-9}
\end{equation}
This agrees with the evolution equation for the displacement of the 
brane, i.e. Eq. (\ref{3-1-19}) at low energies.

\section{Brane fluctuation and anisotropic perturbations
in the bulk}
In this appendix, we show that the brane fluctuation 
interacts with the anisotropic perturbations in the bulk.
We find the anisotropic shear induced by brane fluctuation.
Then the CMB anisotropy induced by this anisotropic shear 
is derived.

\subsection{Anisotropic shear in the brane world}
Let us consider the homogeneous but slightly 
anisotropic 5D spacetime.
The metric for the 5D spacetime is taken as
\begin{equation}
ds^2=e^{2 \gamma(y,t)} dy^2+e^{2 \beta(y,t)} dt^2 + e^{2 \alpha(y,t)}
(\delta_{ij}+ \Pi_{ij}(y,t)) dx^i dx^j,
\label{C-2-1}
\end{equation}
where
\begin{eqnarray}
\Pi_{ij}(y,t) &=& 0, \:\:\:\:\:\:\:\:\: (\mbox{for} \:\:\: i=j), \nonumber\\
         &=& \Pi(y,t) , (\mbox{for} \:\:\: i \neq j ).
\label{C-2-2}
\end{eqnarray}	 
The branes are again located at $y=0$ and $y=l$ respectively.
For a linear anisotropic shear $\Pi(y,t) \ll 1$, the evolution equation for 
$\Pi(y,t)$ is given by
\begin{equation}
-(\Pi''+(3 \alpha'+ \beta'-\gamma') \Pi')+ 
e^{-2 (\beta-\gamma)}(\ddot{\Pi}+(3 \dot{\alpha}-\dot{\beta}+\dot{\gamma})
\dot{\Pi})=0.		
\label{C-2-3}  
\end{equation}
The junction conditions for $\Pi(y,t)$ are given by
\begin{equation}
\Pi_1(t)= \Pi^B_1(t)=0.
\label{C-2-3-1}
\end{equation}
It should be noted that for a linear anisotropic shear $\Pi \ll 1$, 
the evolution of the scale factor of the brane universe is not 
affected by $\Pi$. 
Thus the evolution of the scale factor is still determined locally.
However, the anisotropic shear itself cannot be determined locally.
This can be seen by observing the fact that the evolution equation 
(\ref{C-2-3}) contains $\Pi_2(t)$ if we project the equation 
onto the brane. We cannot determine $\Pi_2(t)$ solely by the 
information on the brane. The solution in the bulk should be
found. Thus solving the bulk evolution equation (\ref{C-2-3})
with the boundary conditions (\ref{C-2-3-1}) is an essential step 
to know the behavior of the anisotropic shear on the brane.

Let us investigate the effect of the brane fluctuation
on the anisotropy of the brane. 
Now suppose that the brane is displaced due to the brane fluctuation. 
We perform the infinitesimal coordinate transformation 
and consider the coordinate system where the brane is located at 
$\hat{y}=0$ in a new coordinate system;
\begin{equation}
x^M \to x^M + \xi^M, \quad \xi^M=(Y(y,t,x^i),T(y,t,x^i),X^i(y,t,x^i)).
\label{C-2-4}
\end{equation}
In an anisotropic spacetime, the coordinate transformation which 
depends on the spatial coordinate is allowed. 
We take the coordinate transformation function as
\begin{equation}
Y =\xi^y(y,t) \omega(x^i), \:\: T=\xi^t(y,t) \omega(x^i), \:\:
X^i = \xi(y,t) \sigma^i(x^i),
\label{C-2-5}
\end{equation}
where $\omega(x^i)$ and $\sigma^i(x^i)$ are some functions of the spatial 
coordinate. The functions $\omega(x^i)$ and $\sigma^i(x^i)$ will be
determined so that the spatial homogeneity of the universe is 
preserved after the coordinate transformation. 
By this coordinate transformation, the metric is transformed as
\begin{eqnarray}
\hat{g}_{00} &=& -e^{2 \hat{\beta}} =
 -e^{2 \beta}(1 + 2 \dot{\xi}^t +2 \dot{\beta} \xi^t+2 \beta' \xi^y), \nonumber\\
\hat{g}_{ij} &=& e^{2 \hat{\alpha}} \delta_{ij} =
e^{2 \alpha} (1+2 \dot{\alpha} \xi^t+2 \alpha' \xi^y)
\delta_{ij},\nonumber\\
\hat{g}_{y y} &=& e^{2 \hat{\gamma}} =e^{2 \gamma}(1+2 \xi^{y'}+
2 \dot{\gamma} \xi^t+2 \gamma' \xi^y),
\nonumber\\
\hat{g}_{0 y} &=& e^{2 \gamma} \dot{\xi}^y -e^{2 \beta} \xi^{t'},
\label{C-1-2}
\end{eqnarray}
where we assumed $\omega(x^i) \sim 1$.
By choosing an appropriate $\xi^t$, we can impose the normal condition 
$\hat{g}_{y0}=0$;
\begin{equation}
\xi^t(y,t)=\int^y_0 dy e^{2(\gamma-\beta)} \dot{\xi}^y(y,t) + T_0(t),
\label{C-1-3}
\end{equation}
where $T_0(t)$ is the residual gauge transformation which depends only on
time $t$.
Then the induced metric on the displaced brane $\hat{A}$ is given by
\begin{equation}
ds^2_{brane A}=-e^{2 \hat{\beta}_0(t)} dt^2 + e^{2 \hat{\alpha}_0(t)}
(\delta_{ij} +\hat{\Pi}_{ij}) dx^i dx^j,
\label{C-1-4}
\end{equation}
where
\begin{eqnarray}
e^{2 \hat{\beta}_0} &=& e^{2 \beta_0}(1 + 2 \dot{T}_0 +2 \dot{\beta}_0 
T_0+2 \beta_1 \xi^y_0), \nonumber\\
e^{2 \hat{\alpha}_0} &=&
e^{2 \alpha_0} (1+2 \dot{\alpha}_0 T_0 + 2 \alpha_1 \xi^y_0).
\label{C-1-5}
\end{eqnarray}

We consider the displacement of the brane induced by brane fluctuation 
Eq. (\ref{5-2-7}) and Eq. (\ref{5-3-3});
\begin{equation}
\xi^y_0=e^{-\gamma_0} f(t).
\end{equation} 
We choose the function $T_0(t)$ so that the lapse function and 
the scale factor are not changed by the transformation;
\begin{equation}
\hat{\alpha}_0=\alpha_0,\quad \hat{\beta}_0=\beta_0.
\label{C-1-6}
\end{equation}
Then $T_0(t)$ is determined by $f(t)$ as 
\begin{eqnarray}
T_0 &=& - \frac{\alpha_1}{\dot{\alpha}_0} e^{-\gamma_0} f(t), \nonumber\\
\dot{T}_0 &=& -\beta_1 e^{-\gamma_0} f(t) - \dot{\beta}_0 T_0.
\label{C-1-7}
\end{eqnarray}
Note that the consistency condition of Eqs.(\ref{C-1-7}) gives
the equation for $f(t)$. 
Using the cosmic time $\tau$ on the brane which is defined by 
$d \tau = e^{\beta_0} dt$, 
the equation for $f(t)$ is givne as  
\begin{equation}
f- \frac{\alpha_{0,\tau}^2}{\alpha_{0,\tau \tau}}
\left(\frac{1}{\alpha_{0,\tau}} f_{, \tau}
- f \right)=0.
\label{C-1-10}
\end{equation}
Here Eq. (\ref{A-4}) and Eq. (\ref{A-0}) were used.
This is nothing but the equation for $f(t)$ derived
in section IV (Eq. (\ref{3-1-20}) with $\zeta_{\ast}=0$). 
Thus the consistency condition is satisfied.

The coordinate transformation given by Eq. (\ref{C-2-5})
induces the additional transformations; 
\begin{eqnarray}
\hat{g}_{y i}&=& e^{2 \gamma} Y_{,i}+ e^{2 \alpha} X^{i '}, \nonumber\\
\hat{g}_{0 i}&=& e^{2 \alpha} \dot{X}^i - e^{2 \beta} T_{,i}, \nonumber\\
\hat{g}_{i j}&=& g_{ij}+ X^i_{,j} +X^j_{,i}.
\label{C-2-6}
\end{eqnarray}
The normal condition $\hat{g}_{yi}=0$ can be satisfied
by choosing an appropriate $X^i$, i.e.
\begin{equation}
X^i(y,t) = - \int^y_0 dy e^{2(\gamma-\alpha)} \xi^y(y,t) \omega(x^i)_{,i}
+X_0(t) \sigma^i(x^i),
\label{C-2-7}
\end{equation}
where $X_0(t)$ is the residual gauge transformation. 
The condition that $\hat{g}_{i0}=0$ on the barne A determines $X_0(t)$ 
as 
\begin{equation}
\dot{X}_0(t) \sigma^i = e^{2(\beta_0-\alpha_0)} T_{0} \omega(x^i)_{,i},
\label{C-2-8}
\end{equation}
where we used Eq. (\ref{C-1-3}).
Now, we take the functions of the spatial coordinate $\omega(x^i)$ and 
$\sigma(x^i)$ in Eq. (\ref{C-2-5}) as 
\begin{eqnarray}
l^2 \omega(x^i)&=& l^2 + (x^1 x^2 +x^2 x^3+x^3 x^1), \nonumber\\ 
l \sigma^1(x^i) &=& x^2+x^3,\quad l \sigma^2(x^i)=x^3+x^1,\quad 
l \sigma^3(x^i)=x^1+x^2. 
\label{C-2-9}
\end{eqnarray}
Then the conditions (\ref{C-2-7}) and (\ref{C-2-8}) become
\begin{eqnarray}
\xi(y,t) &=&  - l^{-2} \int^y_0 dy e^{2(\gamma-\alpha)} \xi^y(y,t)
+ l^{-1} X_0(t), \nonumber\\
\dot{X}_0(t) &=& l^{-1} e^{2(\beta_0-\alpha_0)} T_0(t). 
\label{C-2-10}
\end{eqnarray}
From Eqs. (\ref{C-2-6}), (\ref{C-2-7}) and (\ref{C-2-9}),
we find that the trace part of $g_{ij}$ is not transformed. On the other hand,
the traceless part of $g_{ij}$, namely the anisotropic shear $\Pi(y,t)$, 
is transformed as 
\begin{equation}
\hat{\Pi}(y,t)=\Pi(y,t) - 2 l^{-2} \int^y_0 e^{2(\gamma-\alpha)} \xi^y(y,t) 
+ 2 l^{-1} X_0(t).
\label{C-2-11}
\end{equation}
Then the metric on the brane A after the coordinate transformation is 
given by
\begin{equation}
ds_{brane A}^2=-e^{2 \beta_0(t)} dt^2 +e^{2 \alpha_0(t)}
(\delta_{ij} +\hat{\Pi}_{0ij}(t)) dx^i dx^j,
\label{C-2-12}
\end{equation}
where
\begin{eqnarray}
\hat{\Pi}_{0ij} &=&  0, \:\:\:\: (\mbox{for} \:\:i =j), \nonumber\\
               &=& \hat{\Pi}_0(t)=\Pi_0(t) + 2 l^{-1} X_0(t), 
			  \:\:\:\: (\mbox{for} \:\:\: i \neq j) .
			   \label{C-2-13}
\end{eqnarray}
Here $X_0(t)$ is determined by the brane fluctuation $f(t)$ 
from Eqs. (\ref{C-1-7}) and (\ref{C-2-10}) as
\begin{equation}
\dot{X_0} = -e^{2(\beta_0-\alpha_0)}
				   \frac{\alpha_1 e^{-\gamma_0}}{\dot{\alpha}_0}l^{-1} f(t).
\label{C-2-13-1}
\end{equation}
It should be noted that the brane A remains to be homogeneous after the 
coordinate transformation. 
The junction condition is now imposed in the new coordinate.
Thus the junction condition for $\Pi$ is given by
\begin{equation}
\hat{\Pi}_1=0=\Pi_1 - 2 e^{(\gamma_0-2 \alpha_0)} l^{-2} f(t).
\label{C-2-14}
\end{equation}
The same argument is held for the brane $B$. The junction condition
at the brane B is given by
\begin{equation}
\hat{\Pi}^B_1 =0 =\Pi_1^B - 2 e^{(\gamma_0^B-2 \alpha_0^B)}l^{-2} f^B(t).
\label{C-2-15}
\end{equation}
Thus, now the problem is to solve the wave equation for 
$\Pi(y,t)$, i.e. Eq.(\ref{C-2-3}) with the boundary conditions (\ref{C-2-14}) 
and (\ref{C-2-15}) and to calculate the anisotropic shear 
$\hat{\Pi}_0(t)=\Pi_0(t) + 2 l^{-1} X_0(t)$ on the displaced brane. 
If the solution for $\hat{\Pi}_0$ 
is different from $\Pi_0(t)$, we conclude that 
the brane fluctuation affects the anisotropy of the brane.

\subsection{Anisotropic shear induced by brane flucatuation}
We solve the bulk evolution equation
by the assumption of the nearly static configuration. 
The bulk metric is given by Eq. (\ref{B}). Then
the wave equation for $\Pi$ in the bulk is given by
\begin{equation}
\Pi''-b(t) \frac{4}{l}  \Pi'-b(t)^2 e^{2 b(t)y/l} \left
(\ddot{\Pi}+\left(
\frac{\dot{b(t)}}{b(t)}+3 \dot{\alpha}_0-2 \dot{b(t)}\frac{y}{l}
\right) \dot{\Pi} \right)=0.
\label{C-3-19}
\end{equation}
Here the time dependence of $\Pi$ is assumed to be weaker than the 
$y$-dependence of $\Pi$; 
\begin{equation}
\left( \frac{\partial_t \Pi}{\partial_y \Pi} \right)^2 
= {\cal E} \ll 1.
\label{C-3-20}
\end{equation} 

This wave equation is the same as the wave equation for
$\hat{E}$ (\ref{3-3-19}). Thus we can solve the equation in the same way. 
First let us consider a non perturbed brane. 
The boundary conditions for $\Pi(y,t)$ are given by
\begin{equation}
\Pi_1(t)=\Pi_1^B=0.
\end{equation}
The evolution equation for $\Pi_0$ can be obtained by 
taking the right-hand side of Eq. (\ref{3-3-31}) as zero;
\begin{equation}
\ddot{\Pi}_0+ \left( 3 \dot{\alpha_0} + \dot{b}(t) \frac{e^{-b(t)}}
{\sinh b(t)}\right) \dot{\Pi}_0 =0.
\label{C-3-28}
\end{equation}
We find that the evolution of the anisotropic shear
deviates from the prediction of the 4D Einstein gravity
if $\dot{b} \neq 0$. For $b \to \infty$, the system becomes
effectively one brane system and the 4D Einstein gravity is recovered.

Now let us consider the effect of the brane fluctuation. 
The brane fluctuation acts as a source in the
junction conditions. From Eqs. (\ref{C-2-14}) and 
(\ref{C-2-15}), the junction conditions are given by 
\begin{equation}
\Pi_1=2 b(t) l^{-2} e^{-2 \alpha_0} f(t),\quad 
\Pi_1^B=2 b(t) l^{-2} e^{2b(t)} e^{-2 \alpha_0} f^B(t).
\label{C-3-29}
\end{equation}
We obtain the evolution equation for $\Pi_0$ as
\begin{equation}
\ddot{\Pi}_0+ \left( 3 \dot{\alpha}_0 + \dot{b}(t) \frac{e^{-b(t)}}
{\sinh b(t)}\right) \dot{\Pi}_0 = -\frac{e^{b(t)}}
{\sinh b(t)} e^{-2 \alpha_0} 2 l^{-3} 
(f(t)-e^{-2 b(t)} f^B(t)).
\label{C-3-31}
\end{equation}
The solution for $\Pi_0$ can be obtained as (see Eq. (\ref{3-3-34}))
\begin{equation}
\dot{\Pi}_0= - 2 e^{-3 \alpha_0} 
\left(\frac{e^{b(t)}}{\sinh b(t)} \right) \int 
e^{\alpha_0} l^{-3} (f(t)-e^{-2b(t)} f^B(t)) \: dt.
\label{C-3-34}
\end{equation}
For simplicity we take $f^B(t)=0$. Then using the solution
for $f(t)$ given by Eq. (\ref{5-3-4}) and the scale factor $\alpha_0$
\begin{equation}
f(t) \propto e^{-\frac{(3w+1)}{2} \alpha_0},\quad
e^{\alpha_0} \propto t^{\frac{2}{3(1+w)}},
\label{C-3-35}
\end{equation}
we obtain the solution for $\dot{\Pi}_0$ which depends on 
$f(t)$ as 
\begin{equation}
\dot{\Pi}_0=-2 e^{-2 \alpha_0} \frac{f(t)}{\dot{\alpha}_0}
l^{-3} e^{2 b(t)} {\cal N}(b(t)), \quad {\cal N}(b(t))
=\left(\frac{e^{-b(t)}}{2 \sinh b(t)} \right).
\end{equation}
The observed anisotropic shear $\hat{\Pi}_0$ is given by 
Eq. (\ref{C-2-13}) and Eq. (\ref{C-2-13-1});
\begin{equation}
\dot{\hat{\Pi}}_0=\dot{\Pi}_0+2 e^{-2 \alpha_0}l^{-3} 
\frac{f(t)}{\dot{\alpha}_0}.
\end{equation}
Hence, we obtain the solution for $\dot{\hat{\Pi}}_0$ that depends
on $f(t)$ as
\begin{equation}
\dot{\hat{\Pi}}_0 = -2 e^{-2 \alpha_0} \frac{f(t)}{\dot{\alpha}_0}
l^{-3} {\cal N}(b(t)).
\label{C-3-36}
\end{equation}
Thus one can see that the brane fluctuation becomes a source
of the anisotropic shear. However, if we take $b \to \infty$ 
to consider the one brane model, we get ${\cal N}(b(t)) \to 0$. 
Then the brane fluctuation does not affect the evolution of the 
anisotropic shear. 
Thus the brane fluctuation has no physical degree of the 
freedom in the one brane model.
In two branes model, the brane fluctuation acquire a 
physical degree of freedom through the interaction with
the bulk anisotropic perturbations.

\subsection{Detectability of the brane fluctuation in CMB anisotropy}
In this subsection, we investigate the CMB anisotropy
caused by the anisotropic shear which is induced by the brane fluctuation 
(\ref{C-3-36}). We calculate the CMB anisotropy by following the arguments 
in Ref \cite{Pee}.
The metric of the brane universe is given by
\begin{equation}
ds^2= -dt^2 + e^{2 \alpha_0(t)} (\delta_{ij}+\hat{\Pi}_{0ij}(t)) dx^i dx^j.
\end{equation}
The path of the light ray that reaches our position at $t=t_0$
is given by
\begin{equation}
x^i(t)=\gamma^i x(t), \quad x(t)= \int^{t_0}_t \frac{dt'}
{e^{\alpha_0(t')}},
\label{he}
\end{equation}
where $\delta_{ij} \gamma^i \gamma^j=1$.
Let us consider two comoving observers. The coordinate separation 
of these observers is given by
\begin{equation}
\delta x^i = \gamma^i \delta x.
\end{equation}
The proper distance between two observers at fixed time $t$
is given by 
\begin{equation}
\delta l = \sqrt{g_{ij} \delta x^i \delta x^j}=e^{\alpha_0}
\left(1+ \hat{\Pi}_{0ij} \frac{\gamma^i \gamma^j}{2} \right)
\delta x.
\end{equation}
Then the relative velocity $\delta v$ of two observers is 
\begin{equation}
\frac{\delta v}{\delta l}=
\frac{1}{\delta l} \frac{d \delta l}{d t}=
\dot{\alpha}_0 + \frac{1}{2} \dot{\hat{\Pi}}_{0ij} \gamma^i \gamma^j.
\end{equation}
The difference of the CMB temperature measured by neighbouring 
observers moving apart at speed $\delta v$ is $\delta T/T=
\delta v$. The proper separation $\delta l$ at time $t$ is given 
by $\delta l = \delta t$, so the difference in the CMB 
temperature they measure is 
\begin{equation}
\frac{\delta T}{T} =\left(\dot{\alpha}_0 + \frac{1}{2}
\dot{\hat{\Pi}}_{0ij} \gamma^i \gamma^j \right) \delta t.
\end{equation}
The first term is the usual cooling law in a homogeneous and 
isotropic universe. The observed CMB anisotropy due to the 
anisotropic shear is the result of summing the second term in 
$\delta T/T$ along the path of the photon $x^i(t)=\gamma^i x(t)$;
\begin{equation}
\frac{\Delta T}{T}= \frac{1}{2} \gamma^i \gamma^j 
\int^{t_0}_{t_e} dt \dot{\hat{\Pi}}_{0ij}(t,\gamma^i x(t)).
\end{equation}
From Eq. (\ref{C-2-13}), the anisotropic shear is written as
\begin{equation}
\hat{\Pi}_{0ij}=\hat{\Pi}_0(t) \frac{\partial^2 (l^2 w(x^i))}{\partial x^i \partial x^j},
\quad l^2 w(x^i)=l^2+ x^1 x^2+x^2 x^3+x^3 x^1.
\end{equation}
The integral is evaluated along the path $x^i(t)=\gamma^i x(t)$.
Thus we can consider $\omega$ to be a function of $x(t)$
\begin{equation}
\frac{d \omega}{d x}=\frac{\partial \omega}{\partial x^i}
\frac{dx^i}{d x}=\gamma^i \frac{\partial \omega}{\partial x^i}.
\end{equation}
Hence the temperature fluctuation is written as 
\begin{equation}
\frac{\Delta T}{T}=  \frac{1}{2} \int^{t_0}_{t_e} dt \dot{\hat{\Pi}}_0 
\frac{d^2 (l^2 \omega)}{d x^2}.
\end{equation}
Integrating by parts twice and using $dx=-e^{-\alpha_0} dt$ (Eq. (\ref{he})), 
we get 
\begin{equation}
\frac{\Delta T}{T}= -\frac{1}{2} e^{\alpha_0} \frac{d}{dt} \left[e^{\alpha_0} 
\dot{\hat{\Pi}}_0 \right] l^2 \omega(x^i) -
\frac{1}{2} e^{\alpha_0} \dot{\hat{\Pi}}_0 \gamma^i 
\frac{\partial (l^2 \omega)}{\partial x^i} - 
\frac{1}{2} \int dt \frac{d}{dt} \left[ e^{\alpha_0} \frac{d}{dt}
[e^{\alpha_0} \dot{\hat{\Pi}}_0] \right] l^2 \omega(x^i).
\end{equation}
The first term represents the ordinary Sachs-Wolfe effect. 
Then substituting the solution for the anisotropic shear induced 
by the brane fluctuation (\ref{C-3-36}), the CMB anisotropy 
due to the Sachs-Wolfe effect is given by
\begin{equation}
\frac{\Delta T}{T}= \frac{\dot{{\cal N}}(b(t))}{\dot{\alpha}_0} l^{-1} f(t).
\label{A}
\end{equation}
This result agrees with Eq. (\ref{Q-5}) in the main text. 

\acknowledgements
We would like thank M.Sasaki and J.Soda for helpful comments and 
discussions. This work is supported by JSPS.

\end{document}